\documentclass [review]{elsarticle}

\usepackage{lineno,hyperref}

\journal{Journal of \LaTeX\ Templates}

\usepackage[utf8]{inputenc}
\usepackage{multirow}
\usepackage[table,xcdraw]{xcolor}
\usepackage{graphicx}
\usepackage{amsmath}
\usepackage{amssymb}
\begin{document}


\begin{frontmatter}
\title{High resolution wind-tunnel investigation about the effect of street trees on pollutant concentration and street canyon ventilation}

\author[label1]{Sofia Fellini\corref{mycorrespondingauthor}}
\cortext[mycorrespondingauthor]{Corresponding author}
\ead{sofia.fellini@ec-lyon.fr}
\author[label1]{Massimo Marro}
\author[label3]{Annika Vittoria Del Ponte}
\author[label3]{Marilina Barulli}
\author[label1]{Lionel Soulhac}
\author[label3]{Luca Ridolfi}
\author[label1]{Pietro Salizzoni}

\address[label1]{Univ Lyon, INSA Lyon, CNRS, Ecole Centrale de Lyon, Univ Claude Bernard Lyon 1, LMFA, UMR5509, 69621,
Villeurbanne France}

\address[label3]{Department of Environmental, Land, and Infrastructure Engineering, Politecnico di Torino, Corso Duca degli Abruzzi 24, 10129 Turin, Italy}

\begin{abstract}
Greening cities is a key solution to improve the urban microclimate and mitigate the impact of climate change. However, the effect of tree planting on pollutant dispersion in streets is still a debated topic. To shed light on this issue, we present a wind-tunnel experiment aimed at investigating the effect of trees on street canyon ventilation. An idealized urban district was simulated by an array of blocks, and two rows of model trees were arranged at the sides of a street canyon oriented perpendicularly with respect to the wind direction. Reduced scale trees were chosen to mimic a realistic shape and aerodynamic behaviour. Three different spacings between the trees were considered. A passive scalar was injected from a line source placed at ground level and concentration measurements were performed in the whole canyon. Results show that the presence of trees alters the concentration pattern in the street with a progressive shift from a nearly two-dimensional to a three-dimensional field depending on tree density. Despite the significant change of the concentration field induced by trees, the average level of pollution in the street, and thus the overall ventilation efficiency, does not show a specific trend with the density of trees.

\end{abstract}

\end{frontmatter}

\section{Introduction}

Urban vegetation plays a key role for the livability of cities \cite{bozovic2017blue}. 
Beyond its aesthetic role, the presence of vegetation brings numerous environmental benefits in urban areas. 
Evapotranspiration and shading have cooling effects that mitigate the urban heat island \cite{oliveira2011cooling,georgakis2017determination}. 
The large surface area per unit volume of vegetative structures facilitates particle deposition which acts as a sink for pollutant particles \cite{litschke2008reduction}. 
Moreover, vegetation has a fundamental role in the hydrological cycle: water infiltration in vegetated soils retains stormwater from entering the drainage system, thus reducing the risk of flash floods \cite{livesley2016urban,revelli2022green}.

While the social and environmental benefits of vegetation in cities are well recognized, the role of vegetation on pollutant dispersion is still controversial \cite{janhall2015review} and depends on the non-trivial interaction between the flow field, the vegetative elements, and the surrounding built environment \cite{abhijith2017air}. 
This is particularly true in street canyons, where the turbulent flow field is strongly influenced by the geometry of the canyon, its orientation with respect to the external wind \cite{soulhac2008flow}, the presence of obstacles \cite{buccolieri2022obstacles}, and the properties of the building walls, e.g., wall roughness and temperature \cite{allegrini2013wind,murena2016effect,marucci2019effect,fellini2020street}.
The presence of vegetated fa\c cades on the building walls, for example, alters the near-wall velocity and may result in a reduction of the turbulent air exchange \cite{li2022impact}.
Low-level hedges, instead, generally improve the air quality at street level and thus help reducing the dose of pollutants inhaled by pedestrians \cite{gromke2016influence}.

The interaction between turbulent flow and vegetation is even more complex in tree-lined avenues, where trees occupy a significant volume of the canyon and their effect on pollutant dispersion depends on the properties and shape of the crowns, the height of the trunks, and the planting pattern \cite{vos2013improving,huang2019impacts}.
A pioneering series of wind-tunnel experiments was performed by Gromke et al. \cite{gromke2007influence,gromke2009impact, gromke2012pollutant} to investigate this scenario. 
In their first studies, trees were modelled as a row of small-scale trees with spherical, permeable crowns on thin stems, placed in the middle of a street canyon of unit height to width ratio $H/W$ \cite{gromke2007influence}. The wind was perpendicular to the canyon axis.
The flow field within the canyon and the concentration at the canyon walls were explored by varying different properties of the trees (crown diameter, tree height, tree spacing).
A relevant increase of concentration at the upwind wall and a slight decrease of concentration at the downwind wall were observed.
These variations were more marked when the canyon was occupied by the greatest volume of vegetation (large diameter of crowns and small distance between the trees). 
To better investigate the effect of tree crown porosity, in a later study, trees were replaced with a metallic cage filled with different amounts of synthetic wadding material \cite{gromke2009impact}. 
They found that the concentration at the canyon walls is sensitive to crown porosity only for high porosity values ($>97\%$). 
Adopting the same experimental conditions, Buccolieri et al. \cite{buccolieri2009aerodynamic,buccolieri2011analysis} simulated a large street canyon ($H/W=0.5$) with two rows of trees. 
They also analysed the case of an approaching wind inclined by $45^\circ$ with respect to the street axis.
The aspect ratio of the canyon and the wind direction turned out to be more influential with respect to vegetation density and crown porosity. However, they evidenced that neglecting the presence of vegetation in the streets would lead to significant errors in the predictions of concentration levels.

The modification of the airflow and concentration field within streets due to tree planting has been widely studied also by means of numerical simulations. 
Reynolds-averaged Navier-Stokes (RANS) models proved to be able to qualitatively reproduce the experimental results \cite[e.g.][]{gromke2008dispersion,buccolieri2009aerodynamic,buccolieri2011analysis,gromke2015influence,vranckx2015impact} but Large Eddy Simulations provide a better agreement as they solve the intermittent and unsteady fluctuations of the turbulent flow which plays a major role in ventilation dynamics \cite[e.g.,][]{salim2011numerical,moonen2013performance}.  However, the advantages of LES involve higher computational costs compared to RANS \cite{salim2011numerical}. 
 Merlier et al. \cite{merlier2018lattice} showed that, thanks to its computational efficiency, LES with the lattice Boltzmann method (LBM) is a promising technique to predict dispersion in street canyons with tree plantings.

Despite remarkable advances in numerical models, simulating complex and porous geometries such as trees and their effect on pollutant dispersion still represents a challenge. 
Wind-tunnel experiments are thus highly recommended to improve and validate existing models. 
For these reasons, we present in this work experimental results aimed at evaluating how tree planting influences the concentration field within a street canyon. 
To this aim, we reproduce a street canyon oriented perpendicular with respect to the wind direction, with two lateral rows of trees and a linear source of gas to mimic vehicular emissions. 
To provide a detailed characterization of the scalar field in the three-dimensional street canyon, the concentration is measured on a high-refined measurement grid with around 1000 sampling points for each configuration. 
The reference geometry is that of a two-dimensional canyon closed laterally, i.e. it does not communicate with side streets. 
Although this is an uncommon configuration, it allows us to accurately estimate the exchange rate of pollutants between the canyon and the atmosphere above. 

In Section \ref{sec-exp}, the experimental setup and the adopted measurement techniques are presented. 
In Section \ref{sec-sim}, we discuss the similarity criteria for the aerodynamic modelling of trees and for the boundary  
layer in the wind tunnel. 
The characterization of the concentration field in the street canyon is reported in Section \ref{sec-ventil}, together with the estimate of the ventilation efficiency. 
Moreover, in Section \ref{sec-extrem}, we investigate the effect of trees on pollutant concentration in streets with different geometrical boundaries (street intersections, open or closed canyon).
Finally, the conclusions and perspectives of the work are presented in Section \ref{sec-concl}.

\section{Experimental setup and measurement techniques}\label{sec-exp}

\subsection{Wind-tunnel setup}\label{sec-wtsetup}
The experiments were performed in the atmospheric wind tunnel at the Laboratoire de M\'ecanique des Fluides et d'Acoustique (LMFA) at the \'Ecole Centrale de Lyon. 
The aerodynamic circuit is composed by an axial fan which induces wind velocities between 0.5 and 6 m/s, flow diverging and converging systems, and an upwind grid for the generation of homogeneous turbulence. 
A heat exchanger system regulates the air temperature with a precision of $0.5$ K. 
The test section of the wind tunnel is 12 m long, 3.5 m wide, and 2 m high. 

\begin{figure}
\centering
\includegraphics[width=1\columnwidth,trim=0 0 1  0,clip]{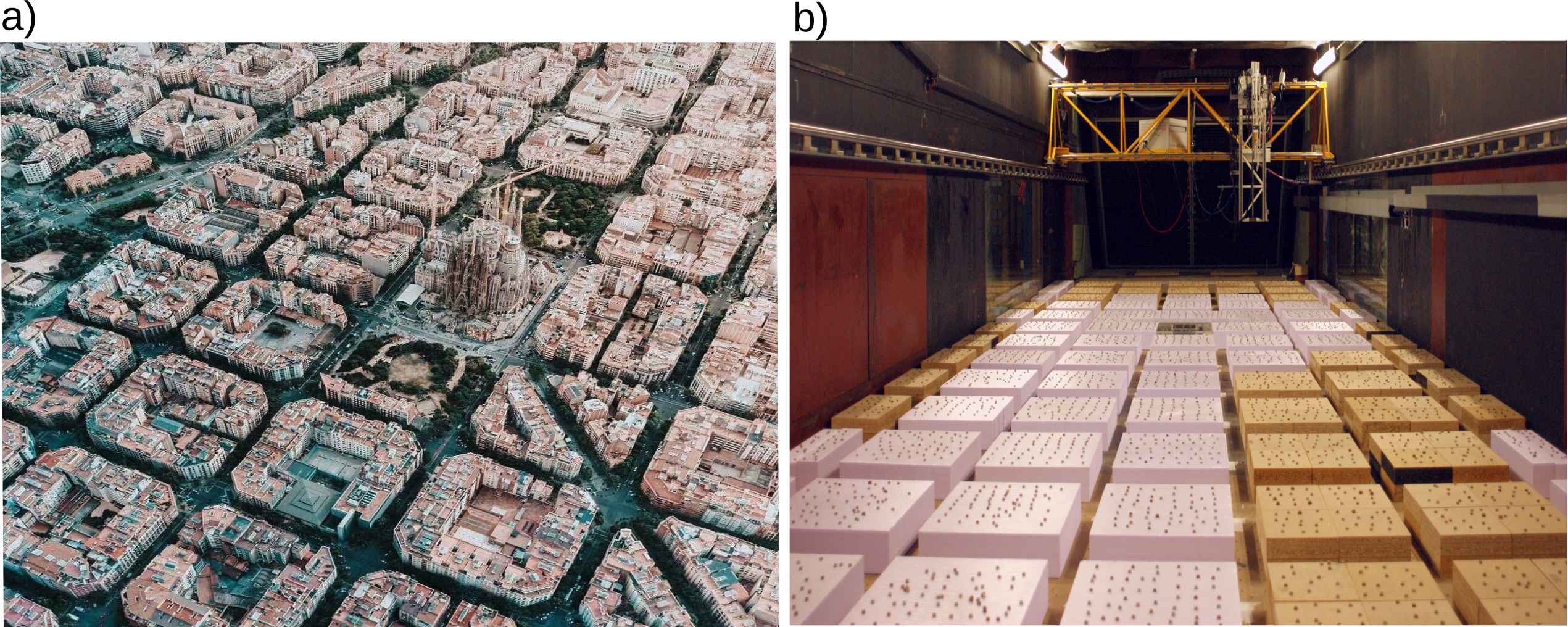}
\caption{(a) Aerial view of the city of Barcelona (Spain). Source: \textit{Barcelona From Above} by Ian Harper. (b) The modelled urban canopy in the wind tunnel.}\label{fig_canopy}
\end{figure}

\begin{figure}
\centering
\includegraphics[width=1\columnwidth,trim=0 0 1  0,clip]{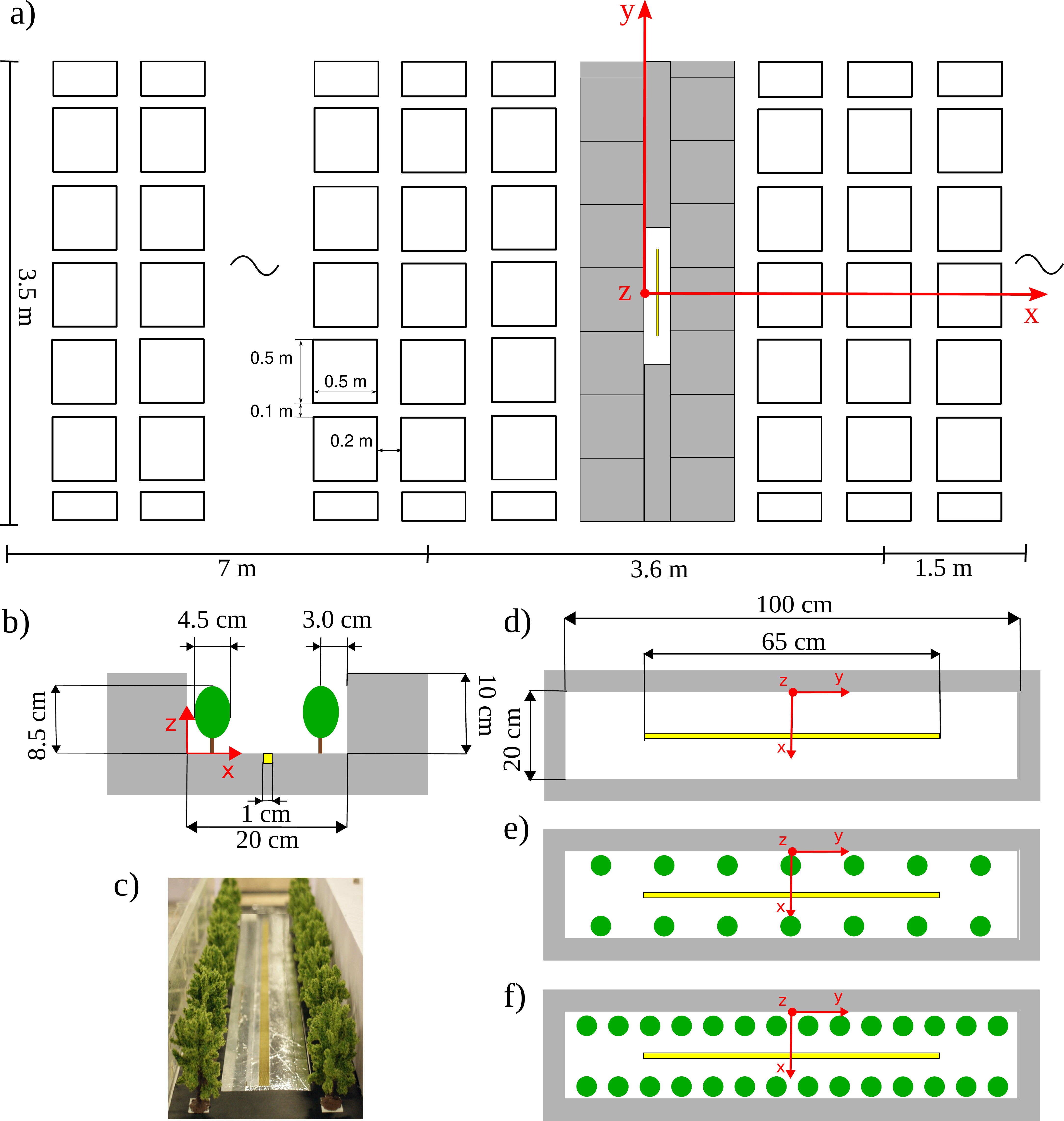}
\caption{(a) Sketch of the urban canopy in the test section of the tunnel. The blocks delimiting the reference street canyon are coloured grey. Sketch (b) and photo (c) of the front view of the street. 
Top view of the street canyon model for the different configurations of tree density: (d) \textit{Zero}, (e) \textit{Half}, (f) \textit{Full}. The yellow line represents the pollutant source}\label{fig_setup}
\end{figure}

To simulate an idealized urban district (Fig. \ref{fig_canopy}) and the turbulent flow within and above it, the floor of the entire test section was overlaid with an array of square blocks (panel a in Fig. \ref{fig_setup}). 
The blocks were 50 cm wide and 10 cm high and made of wood and polystyrene. 
The spacing between the obstacles was 10 cm in the spanwise direction and 20 cm in the streamwise direction.  
In this way, we obtained a street network composed of square canyons (height to width ratio $H/W=$1) aligned with the wind direction intersecting larger perpendicular streets ($H/W=0.5$).  
The two different proportions were selected to avoid channeling effects along the wind direction and to recreate tree-lined boulevards in the direction perpendicular to the wind.
The blockage ratio of the model to the cross-section of the wind tunnel was $5\%$.

A neutrally stratified boundary layer approximately 1.1 m depth was generated by combining the effect of a row of 0.95 m high Irwin spires \cite{Irwin1981}, placed at the beginning of the test section, and the building-like obstacles on the floor. 
Moreover, the obstacles were covered by 5 mm high nuts to generate further roughness and accelerate the full development of the boundary layer. 
The free stream velocity at the top of the boundary layer ($U_\infty$) was kept constant at 5 m/s. More details about the boundary layer above the obstacles is given in Section \ref{sec-boundary}.

The reference street canyon (see photo reported in Fig. \ref{fig_setup}.c) was placed perpendicular to the wind direction, at a distance of approximately 9 meters from the beginning of the test section. Its length (L), width (W), and height (H) measured 1.0 m, 0.2 m, and 0.1 m, respectively (Fig. \ref{fig_setup}.b and d), thus providing the following aspect ratios: $H/W=0.5$, $L/H=10$, and $L/W=5$.
In a 1:200 scale, the street canyon matches fairly well a typical tree-lined boulevard, 40 m wide and flanked by 20 m high buildings, as in typical European city centres (e.g., Barcelona, Turin, Lyon) \cite{soulhac2010dispersion}. 
Note that, in the experiment, the length of the canyon does not cover the whole width of the test section as it is closed laterally by polystyrene blocks. This makes the canyon similar to the one investigated numerically by Moonen et al. \cite{moonen2011evaluation}.
This geometry was chosen in order to define a control volume to compute the mass balance for an accurate estimate of the ventilation efficiency (see Section \ref{sec-ud}). 
We are aware that other boundary conditions are possible, for instance street intersections or an indefinitely long canyon. 
The effects of different lateral boundaries will be discussed in Section \ref{sec-extrem}.

To simulate urban vegetation in the street canyon, model trees were aligned along two lateral rows 14 cm apart. 
Three different configurations for the tree density were analysed: in configuration \textit{Zero} (panel d), the street canyon was empty. 
In configuration \textit{Half} (panel e), seven trees equally spaced by 14 cm (9.5 cm distance between the crowns) were arranged along each lateral row. 
In configuration \textit{Full} (panel f), the lateral rows were composed of fourteen trees equally spaced by 7 cm (2.5 cm distance between the crowns). 
The aerodynamic characterization of model trees is reported in Section \ref{sec-trees}.

To simulate vehicular emissions, a tracer was emitted by a linear source at the centre of the reference street canyon. 
The source is the same adopted by Marro et al. \cite{marro2020high}, to which we refer for more details. 
It consists of a stainless steel tube pierced with needles emitting a gas mixture in a homogenization chamber, located in a slot cut in the tunnel floor. 
From this chamber, the gas is released at street level from a 65 cm long and 1 cm wide metallic grid. 
The small diameter of the needles produces substantial pressure drop thus rendering the gas release insensitive to local pressure fluctuations in the street canyon. 
This design aims at minimizing the vertical momentum and maximizing the lateral homogeneity of the emission, following the indications of Meroney et al. \cite{meroney1996study}.
The source released a mixture of air and ethane. 
The latter is used as a passive tracer, since it is detectable by the Flame Ionisation Detector (FID) (see Section \ref{sec-meas-tech}) and has approximately the same density of air (the density ratio between ethane and air being about 1.03). 
The total air-ethane flow rate was 4.0 l/min, with a percentage of ethane of $5\%$ in volume (0.2 l/min). 
The injection velocity at street level (estimated as the ratio between the total flow rate and the release surface at street level) was around 0.01 m/s.
The supply of the two gases was regulated by two digital mass-flow controllers (Alicat Scientific MC-Series) working in a range between 0.1 and 20 Nl/min (for air) and 0.01 and 2 Nl/min (for ethane), with an accuracy of 0.5\%.

\subsection{Measurement techniques}\label{sec-meas-tech}

The concentration field within the reference street canyon was measured using a Flame Ionisation Detector (HFR400 Fast FID) \cite{fackrell1980flame}, which is commonly used for measurements in urban-like geometries \cite{pavageau1999wind,carpentieri2012wind,fellini2020street}.  
A straight 30 cm long sampling capillary tube, with radius $1.27\times10^{-4}$ m, was mounted on the Flame Ionisation Detector head, which was positioned above the test section so as not to affect the flow field. 
The imposed pressure drop along the capillary tube was 33330.6 Pa.
The sampling frequency of the FID signal was fixed at 1000 Hz \cite{nironi2015dispersion}. The instrument works in the range 0-10 Volt and it can detect concentration values between 0 and 5000 ppm with an accuracy of about 1-2 ppm. A more detailed description of the FID is provided by Marro et al. \cite{marro2020high}.

The measurements were performed in statistically steady conditions: a constant flow rate of ethane was injected from the ground level source and the concentration within the cavity was measured at around 1000 sampling points for each configuration of tree density. 
The measurement grid was defined to characterize the entire three-dimensional volume in detail (see Fig. S2 in the Supplementary Material). 
For each point, we fixed a sampling time of 2 minutes which provides a reliable estimate of the mean concentration. Moreover, before and after each acquisition, the background concentration was recorded by stopping the emission for 15 seconds (and leaving time for the transients to settle).
The background concentration, which was assumed to evolve linearly with time from its initial to its final value, was then subtracted from the signals \cite{marro2020high,vidali2022wind}.

The velocity field above the obstacles was characterized (Section \ref{sec-boundary}) by means  of a Hot-Wire Anemometer (HWA) at constant temperature, using an X-wire probe with acceptance angle of $45^\circ$.
In this way, two velocity components of the velocity field were measured simultaneously.  
The platinum probe wire was 1 mm long and with a diameter of 5 $\mu$m. 
The small size of the hot-wire element enables good spatial resolution of the velocity field while the low thermal inertia of the material ensures fast response, allowing the detection of high-frequency fluctuations of the turbulent flow \cite{comte1976hot}.
An acquisition time of 1 minute at a frequency of 4000 Hz was adopted for each sampling point. Calibration of the HWA was carried out using a Pitot tube to measure a reference velocity.\\

For the aerodynamic characterization of the model trees (presented in Section \ref{sec-trees}), their drag coefficient was measured in a small closed-circuit wind tunnel with a 30 cm $\times$ 30 cm test section being able to generate velocities up to 25 m/s. 
The tunnel was equipped with an external load cell with a precision of 0.01 N.
Different layouts of trees were attached to a removable plate connected with the load cell. 
The drag coefficient was estimated for a varying wind velocity inside the tunnel.
Moreover, the aerodynamic porosity of the model trees was evaluated by performing velocity measurements upwind and downwind a single tree on a regular grid of points. 
Since in this case we needed to measure only the average velocity, we used a Pitot tube.


\section{Similarity criteria}\label{sec-sim}

\subsection{Aerodynamic characterization of model trees}\label{sec-trees}

To investigate the effect of trees in urban areas by means of wind-tunnel experiments, buildings and vegetative structures need to be modelled in small scale. 
Similarity criteria are then necessary to transfer small-scale findings in the wind tunnel to full-scale applications. 
For impermeable and rigid structures, like buildings, dynamical similarity between the experiment and the real application exists if the model and the full-scale object are geometrically similar and the value of the Reynolds number is the same \cite{tritton2012physical}. 
In case of fully turbulent flows around bluff bodies with sharp edges, this condition is weakened and flow similarity is assumed as far as the Reynolds number is sufficiently large.
On the other hand, less knowledge is available about the appropriate similarity criteria for vegetative structures.
From a fluid dynamical point of view, vegetation is a complex porous medium made of branches and leaves giving rise to the development of boundary layers, wakes, and recirculation zones \cite{gromke2008aerodynamic}.
Moreover, due to their flexibility, trees can sway with the wind and induce fluid-structure interactions. 

In previous wind-tunnel experiments, trees, windbreaks, and canopies have been modelled by using different materials, e.g., brushes, cotton balls, metal screens, and plastic stripes. 
Aerodynamic validation of the adopted structures was done by analysing different fundamental features of the interaction between the trees and the flow field, as the drag coefficient, the characteristics of the wakes behind the trees \cite{meroney1968characteristics}, the ratio between tree height and roughness length \cite{meroney1980wind}, the leaf area density \cite{chen1995wind}, or the sway frequency \cite{stacey1994wind}.
Gromke and Ruck \cite{gromke2009impact,gromke2012pollutant} used the pore volume fraction and the pressure loss coefficient to characterize the porous media that mimicked avenue trees. Gromke \cite{gromke2011vegetation} also established a similarity criterion based on the pressure loss coefficient. 
More systematically, Gromke and Ruck \cite{gromke2008aerodynamic} analysed the aerodynamic characteristics of 12 small-scale modelled trees made of different materials and porosity. 
Measurements of the drag coefficient and of the flow field around the crowns evidenced the drag coefficient as a key scale parameter for the modelling of trees. 
Manickathan et al. \cite{manickathan2018comparative} compared the aerodynamic behaviour of model and natural trees in a wind tunnel. 
They found that, together with the drag coefficient, the aerodynamic porosity of the tree crown is another key parameter to compare natural and model trees. 

In accordance with these studies, we mimicked natural trees with plastic trees for railway modelling and we characterized their aerodynamic behaviour by estimating their aerodynamic porosity and their drag coefficient. 
We also investigated their optical porosity.  
The trees were 8.5 cm high and 4.5 cm wide, with crowns in plastic porous material on plastic trunks. Under the conditions of the experiment, tree models behaved like rigid bodies and thus deformations and fluid-structure interactions could be neglected.   \\

\begin{figure}
\centering
\includegraphics[width=1\columnwidth]{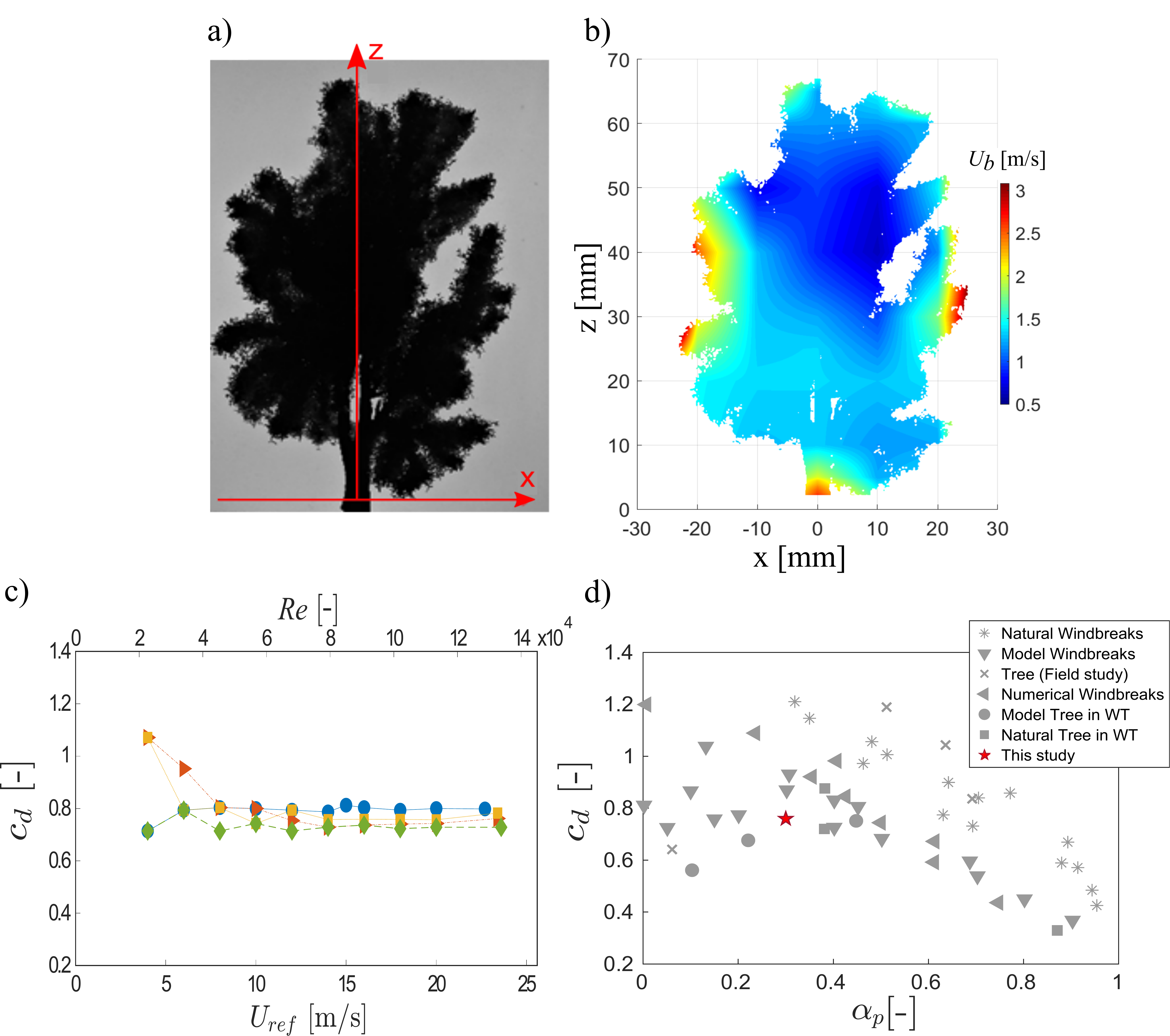}\\
\caption{a) Front photo of the model tree. b) 2D velocity field downstream the tree obtained by spatial interpolation of point velocity measurements for the estimate of $\alpha_p$. c) Drag coefficient as a function of the Reynolds number and the wind velocity for four different alignments (reported by different colours) of a single model tree, obtained by rotating the vertical axis of the tree at $90^\circ$ intervals. d) Drag coefficient and aerodynamic porosity for various model trees and natural trees. From the study of Manickathan et al. \cite{manickathan2018comparative}.}
\label{cd_porosity}
\end{figure}

Aerodynamic porosity ($\alpha_p$) is defined \cite{guan2003wind} as the ratio of the time average wind speed behind the obstacle ($U_b$) and the average speed of the approaching wind ($U_{ref}$): 
\begin{equation}\label{aero_porosity}
\alpha_p=\frac{\int_{A_c} U_b(x,y) dA_c}{\int_{A_c} U_{ref}(x,y) dA_c},
\end{equation}
where $A_c$ is the projected frontal area of the obstacle. 
In other words, aerodynamic porosity determines the portion of the flow that passes through the porous material with respect to the flow that diverges from the obstacle. 
We note that this aerodynamic parameter was initially introduced to characterize windbreaks. However, its application has been more recently extended to individual trees \cite{bitog2011wind,lee2014shelter,manickathan2018comparative}.
To estimate $\alpha_p$, we performed velocity measurements upstream and downstream a single tree that was placed in a homogeneous flow where (i.e. $U_{ref}\approx$const), so that the integral at the denominator in Eq. \ref{aero_porosity} becomes $U_{ref} \cdot A_c$.
Following the indication of Guan et al. \cite{guan2003wind}, the velocity behind the tree was measured in the first plane not occupied by the tree branches. 
Velocities were measured on a regular and dense grid and a two-dimensional velocity field was obtained through spatial interpolation. 
The spatial average velocity was then estimated by integrating the velocity field over the tree silhouette (Fig. \ref{cd_porosity}.b). 
The mean speed upstream ($U_{ref}$) and downstream ($U_b$) the tree were 4.95 m/s and 1.48 m/s, respectively. 
By means of Eq. \ref{aero_porosity}, we obtained $\alpha_p=0.3$, a value in line with that of common natural trees, as hollies and cypresses (see square markers in Fig. \ref{cd_porosity}.d from the study of Manickathan et al. \cite{manickathan2018comparative}). \\

The optical porosity $\beta_p$ is another commonly used parameter to characterize the vegetation and it can be easily estimated by elaborating digital photos \cite{velardeestimation}. 
It is defined as the ratio between the open surface of a porous material and its total surface. 
Through a digital elaboration of the photo capturing the frontal view of the model tree (Fig. \ref{cd_porosity}.a), we delimited the silhouette of the tree and obtained its cross-section, $A_c\approx3.5 \times 10^{-3}$ m$^2$. 
We then estimated the optical porosity, $\beta_p\approx0.05$, as the ratio of the number of white pixels to the total number of pixels within the silhouette of the tree. 
According to the empirical relationship found experimentally by Guan et al. \cite{guan2003wind}, the optical porosity ($\beta_p$) is related to the aerodynamic porosity as: 
\begin{equation}\label{relation_porosity}
\alpha_p\simeq\beta_p^n,
\end{equation}
where the exponent $n$ was estimated by Guan et al.  \cite{guan2003wind} equal to 0.4 for realistic windbreak. 
Introducing our estimated values for $\alpha_p$ and $\beta_p$ in Eq. \ref{relation_porosity}, we find $n\approx$ 0.402 that is consistent with the cited study.
The relation in Eq. \ref{relation_porosity} can then be conveniently used for deriving the aerodynamic porosity when velocity measurements cannot be directly performed.\\ 

The drag coefficient is defined as:
\begin{equation}\label{drag_definition}
c_d=\frac{2F}{\rho_a U_{ref}^2A_c},
\end{equation}
where $F$ is the drag force [N], $\rho_a$ is air density ($\approx 1.18$ kg/m$^{3}$ at $25^\circ$C), $U_{ref}$ is the reference velocity [m/s] for the approaching wind, and $A_c$ is the projected frontal area of the tree [m$^2$]. 
The drag force $F$ was measured by means of a load cell, while the velocity $U_{ref}$ was measured with a Pitot tube, as explained in Section \ref{sec-meas-tech}.

In Fig. \ref{cd_porosity}.c, we report the drag coefficient as a function of the wind velocity $U_{ref}$ and Reynolds number $Re$ for four different faces of a single model tree, obtained by rotating the vertical axis of the tree at intervals of $90^\circ$.
The Reynolds number was calculated as $Re=U_{ref}H_T/\nu$, where $H_T$ is the tree height and $\nu$ is the air kinematic viscosity ($\nu \approx 1.55 \times 10^{-5}$ m$^2$/s at a temperature of 25$ ^\circ$C).
Except for the values at low speed (where the experimental uncertainty of the measurement is large), the drag coefficient rapidly converges to a constant value around 0.75. 
As shown in Fig. \ref{cd_porosity}.d, this value is in line with the drag coefficient of natural trees and confirms that the model trees adopted in this study present realistic aerodynamic properties. 

We note that natural trees undergo foliage reconfiguration and their drag coefficient decays with increasing wind speed \cite{manickathan2018comparative}. 
This is not found in our model trees that do not deform.
However, since in this study we focus on moderate velocities in a street canyon, we are not interested in reproducing the flexibility of natural trees.

\subsection{Characterization of the boundary layer}\label{sec-boundary}

The turbulent boundary layer developing above the idealised urban district was characterized by measuring and focusing on the statistical properties of the turbulent flow. To this aim, we measured vertical velocity profiles in different positions of the wind tunnel. 

\begin{figure}
\centering
\includegraphics[width=1\columnwidth]{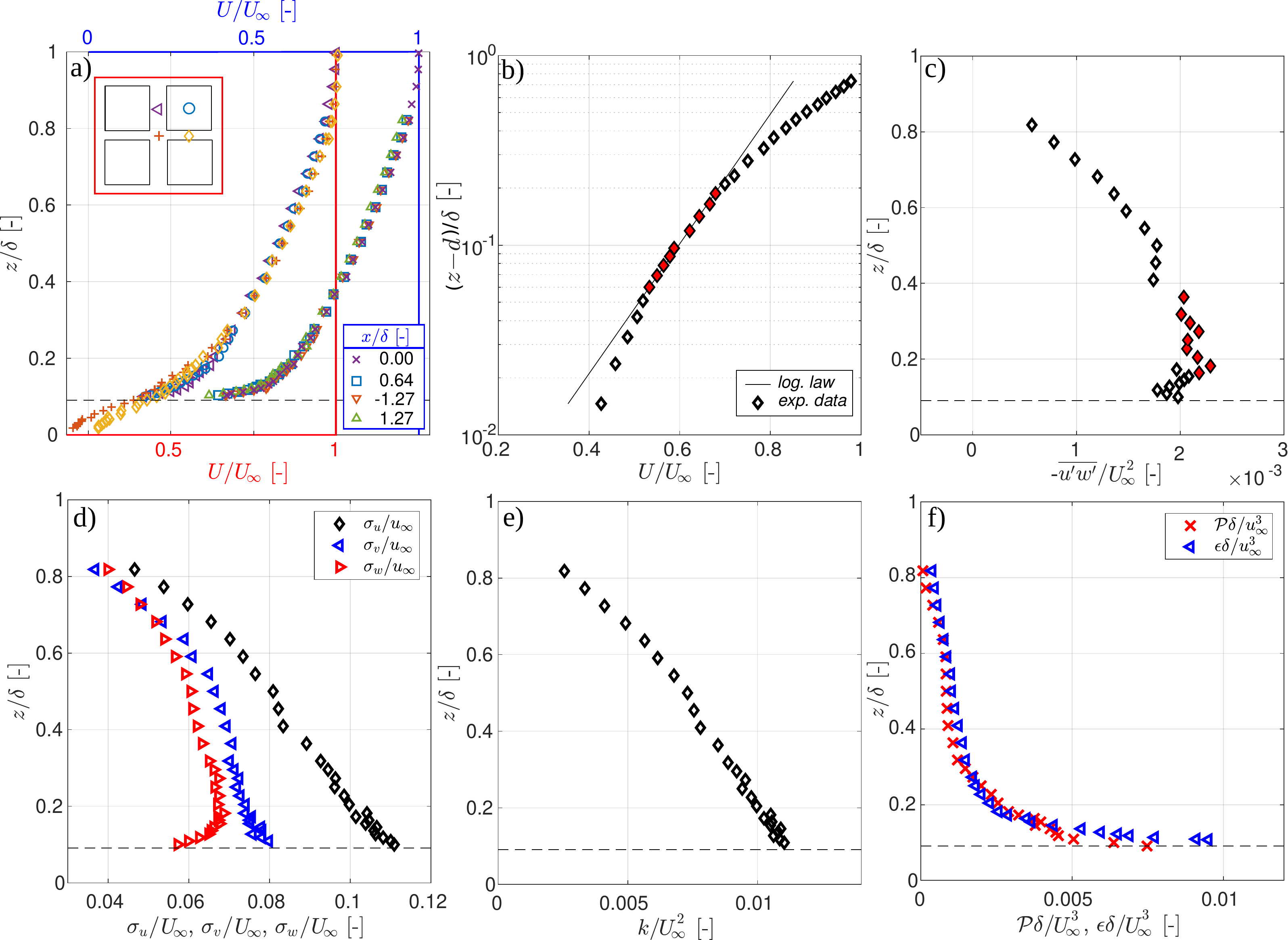}
\caption{a) Mean velocity at 4 different position in a spatial periodic unit (red bottom $x$-axis) and at 4 different distances along the streamwise direction of the wind tunnel (blue top $x$-axis). For the two groups of profiles, a vertical line corresponding the $U/U_\infty=1$ is reported. The horizontal dashed line corresponds to the canyon roof level ($H$). b) Mean velocity obtained as average over four different positions. The line represents the logarithmic law with $u_*/U_\infty=0.051$, $z_0/\delta=9\times10^{-4}$, and $d/\delta=0.085$. The full symbols indicate the region where the logarithmic law applies. 
c) Reynolds stresses $-\overline{u'w'}$. The full symbols indicate the constant-stress region. d) Standard deviation of the three velocity components. e) Turbulent kinetic energy. f) Production and dissipation rate of turbulent kinetic energy.}\label{boundary_layer}
\end{figure}

The evolution of the boundary layer along the central axis of the wind tunnel (from $x=-1.27$ m to $x=1.27$ m) is shown in the right part of Fig. \ref{boundary_layer}.a (blue top $x$-axis). 
The good overlapping between the curves reveals that the flow is fully developed when it approaches the reference canyon ($x=0$), i.e. its development in the stream-wise direction is so slow that changes over the fetch can be neglected. 
As already introduced in Section \ref{sec-wtsetup}, the free-stream velocity ($U_\infty$) is around 5 m/s while the height of the boundary layer ($\delta$) is around 1.1 m. 
The ratio $\delta/H$=11 is in the typical range ($\delta/H\approx 10-20$) for atmospheric flow over urban canopies \cite{perret2019atmospheric}.
The characteristic Reynolds numbers based on the obstacle height are $Re_\infty=U_\infty H/\nu \approx 3.3\times 10^4$ and $Re_H=U_H H/\nu \approx 1.25\times 10^4$, where $U_H=1.94$ m/s is the mean horizontal velocity at $z=H$. 
These values are sufficiently high to ensure fully-developed turbulent flow.
For a square cavity, Allegrini et al. \cite{allegrini2013wind} obtained a Reynolds independent flow for $Re_\infty > 1.3\times 10^4$,  while Castro and Robins \cite{castro1977flow} and Marucci and Carpentieri \cite{marucci2019effect} showed that the condition was met for $Re_H>4000$.


Four velocity vertical profiles were measured at different positions within a periodic unit of the urban canopy (see the inset and the velocity profiles on the left side of Fig. \ref{boundary_layer}.a, red bottom $x$-axis). 
The influence of the individual obstacles is evident in the lower part of the velocity profiles where the velocity profiles collected in different positions show discrepancies one from another up to $z=0.15\,\delta$ ($z=1.65\,\delta$). 
This height can be considered as the upper limit of the so-called roughness sublayer \cite{oke2017urban} and is in line with respect to the typical value of 2$H$ estimated in European cities with regular building geometries \cite{fisher2006meteorology}. 
Above this height, the inertial sublayer develops and the flow variables depend on the vertical coordinate only. 
In this zone, the mean velocity profile is usually modelled by the logarithmic law (Fig. \ref{boundary_layer}.b):
\begin{equation}
\frac{U}{u^*}=\frac{1}{\kappa} \ln \frac{z-d}{z_0},
\end{equation} 
where $\kappa=0.4$ is the Von K\'arm\'an constant, $z_0$ is the aerodynamic roughness length, $d$ is the zero-plane displacement, and $u_*$ is the friction velocity.
In the literature, several techniques have been developed to determine the values of these parameters \cite{raupach2006momentum}. 
In the Supplementary Material (Section S1), we compare the results from two different methods and we obtain the following values: $u^*=0.29$ m/s ($u_*/U_\infty=0.051$), $d=0.094$ m ($d/\delta=0.085$), $z_0=1\times10^{-3}$ m ($z_0/\delta=9\times10^{-4}$). 
The normalised roughness length $z_0/H$ and zero-plane displacement $d/H$ are in the reasonable range of variation proposed by Grimmond and Oke \cite{grimmond1999aerodynamic}, who collected and reviewed different estimate methods based on the packing density of buildings, i.e. on the non-dimensional plan area $\lambda_p$ (here $\lambda_p=0.6$) and the non-dimensional frontal area $\lambda_f$ (here $\lambda_f=0.12$).\\

Figure \ref{boundary_layer}.c reports the vertical profile of the Reynolds stresses, obtained as a spatial average over the four horizontal positions reported in the inset of Fig. \ref{boundary_layer}.a. The constant-stress region is represented by full markers. 
The vertical profiles of the standard deviation of the velocity components ($\sigma_u$, $\sigma_v$, and $\sigma_w$) and of the turbulent kinetic energy ($k$) are reported in Figs. \ref{boundary_layer}.d and e, and are in line with previous experiments \cite{garbero2010experimental} performed in the same wind tunnel. 
The turbulent velocity components above roofs, scaled on the friction velocity $u_*$, provide $\sigma_u/u_*\approx2.1$, $\sigma_v/u_*\approx1.6$, and $\sigma_w/u_*\approx1.3$. 
It is worth noting that these values are consistent with those provided by Britter and Hanna \cite{britter2003flow} for typical urban areas ($\sigma_u/u_*=2.4$, $\sigma_v/u_*=1.9$, and $\sigma_w/u_*=1.3$), albeit the horizontal and lateral components are slightly underestimated.
We also provide in Fig. \ref{boundary_layer}.f the vertical profile of the turbulent kinetic energy dissipation rate ($\varepsilon$), which is a fundamental parameter for turbulence closure models in CFD simulations.
The dissipation rate was estimated from the HWA measurements as:  
\begin{equation}
\varepsilon=\frac{15\nu}{U^2}\overline{\left( \frac{\partial u}{\partial t}\right)^2},
\end{equation}
by employing the isotropic approximation and Taylor's hypotheses of frozen turbulence \cite{hinze1975turbulence}.
The vertical profile of $\varepsilon$ agrees well with the production rate of turbulent kinetic energy, here estimated as $\mathcal{P}\approx -\overline{u'w'} \frac{\partial U}{\partial z}$. This shows that, in most of the boundary layer, the production and dissipation rate of turbulent kinetic energy are in local equilibrium \cite{nironi2015dispersion}, a condition required to recover a typical logarithmic boundary layer profile \cite{tennekes1972first}.\\

\section{Street canyon ventilation}\label{sec-ventil}

\subsection{Mean concentration field}\label{sec-conc}

The mean concentration field inside the street canyon was characterized for the three configurations of tree density presented in Fig. \ref{fig_setup}. 
As mentioned above (Section \ref{sec-meas-tech}), the concentration of ethane - released from the line source - was measured, by means of a Flame Ionization Detector, at around 1000 sampling points (for each configuration) distributed on a three-dimensional grid. 
The non-dimensional concentration is expressed as $\overline{C^*}=CU_\infty L_s \delta/Q_{et}$, where $C$ is the time-averaged concentration of ethane in each sampling point, $L_s$ is the source length, and $Q_{et}$ is the mass flow rate of ethane. 
In the following, the results are presented in two-dimensional sections obtained from linear interpolation of the measured data. 
For a complete visualisation of the concentration field inside the canyon refer to Section S3 of the Supplementary Material.

Figure \ref{fig_conc_xz} shows two cross-sections for each configuration of tree density. 
Panels a-c report the concentration field on a lateral cross-section ($y/H\approx$-1.5), whereas panels d-f correspond to the central (around $y/H\approx$0) cross-section. 
Regardless of the presence of trees, a clear increase in the concentration from the downwind wall to the upwind wall can be observed in all the sections. 
This pattern is in accordance with previous studies \cite{gromke2007influence,gromke2009impact} and evidences the action of the main recirculating cell of the velocity field inside the canyon: 
fresh air enters the canyon at the downwind wall and transports the pollutant (emitted in the centre of the street) to the upwind wall, where part accumulates at the lower corner, part is moved outside, and part is entrained towards the downwind wall. 
The inhomogeneity of the concentration field along the $x$-axis results in a significant difference in air quality in the lower part of the canyon (i.e. at $z/H=0.2$). 
This difference is accentuated when trees are present: in a canyon without vegetation (Fig. \ref{fig_conc_xz}.a and d), pollutant concentration at the downwind wall ($x/H=0.1$ and $z/H=0.2$) is roughly 3 times lower than the one at the upwind wall ($x/H=0.9$ and $z/H=0.2$), while in presence of trees this difference along $x$ increases up to 8 times in the lateral section (Fig. \ref{fig_conc_xz}.b and c). 
Although the lowest measured horizontal section ($z/H=0.2$, corresponding to 4 m at the 1:200 scale) is too high to represent the pedestrian level (which is usually considered to be at 1.5 m), the concentration field at this height highlights the strong impact of trees on concentration spatial pattern (see Fig. S6 in the Supporting Information for further details). 
We also remark that, for the non-vegetated canyon, the concentration field remains almost unchanged along the $y$-axis (panels \ref{fig_conc_xz}.a,d) while the presence of trees alters this behaviour: in both the \textit{Half} and \textit{Full} configurations, pollutant concentration in the central section (panels \ref{fig_conc_xz}.e,f) is significantly lower than in the lateral one (panels \ref{fig_conc_xz}.b,c).

\begin{figure}
\centering
\hspace*{-0.5cm}
\includegraphics[width=1.2\columnwidth]{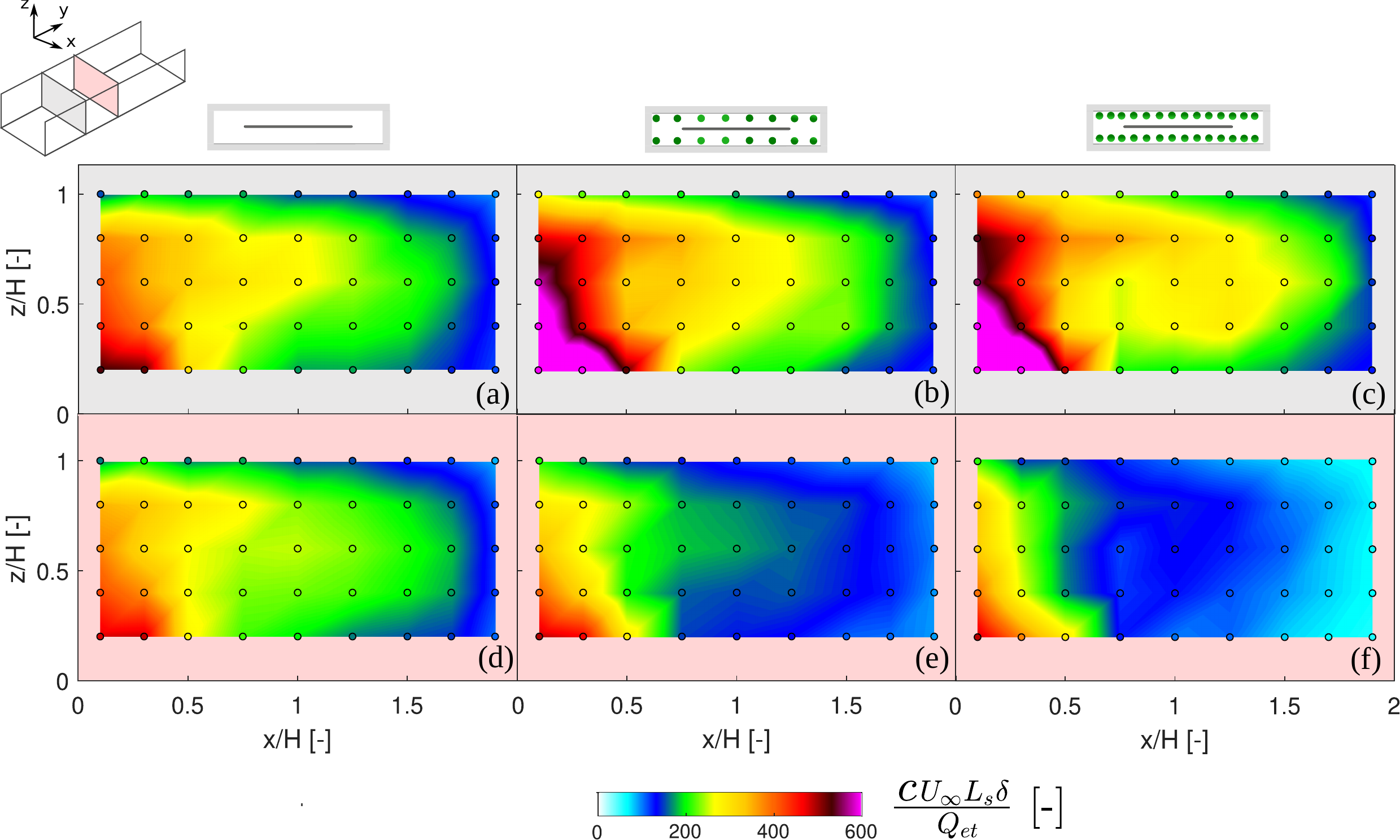}
\caption{Mean concentration of the passive scalar on a lateral cross-section at $y/H\approx-15$ (first line) and in the centre of the canyon $y/H\approx-0$ (second line). \textit{Zero} (a and d), \textit{Half} (b and e) and \textit{Full} (c and f) configurations are shown. Measurement points are reported as circles colored according to the measured value.} \label{fig_conc_xz}
\end{figure}

\begin{figure}
\centering
\includegraphics[width=1\columnwidth]{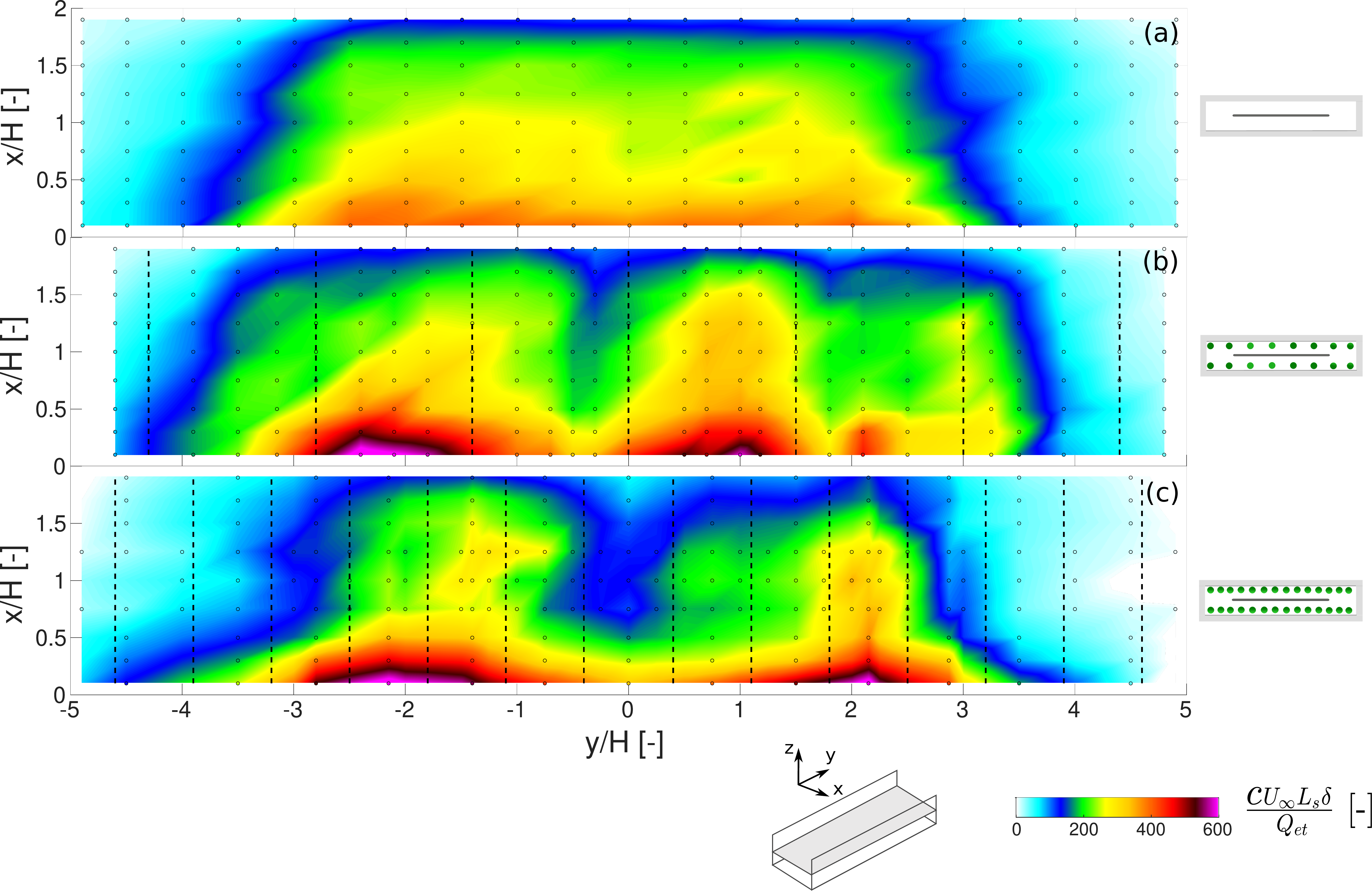}
\caption{Mean concentration of the passive scalar on the horizontal section at $z/H$=0.6. \textit{Zero} (a and d), \textit{Half} (b and e) and \textit{Full} (c and f) configurations are shown. The position of trees is represented by dashed lines.  Measurement points are reported as circles colored according to the measured value.} \label{fig_conc_yx}
\end{figure}

\begin{figure}
\centering
\includegraphics[width=1\columnwidth]{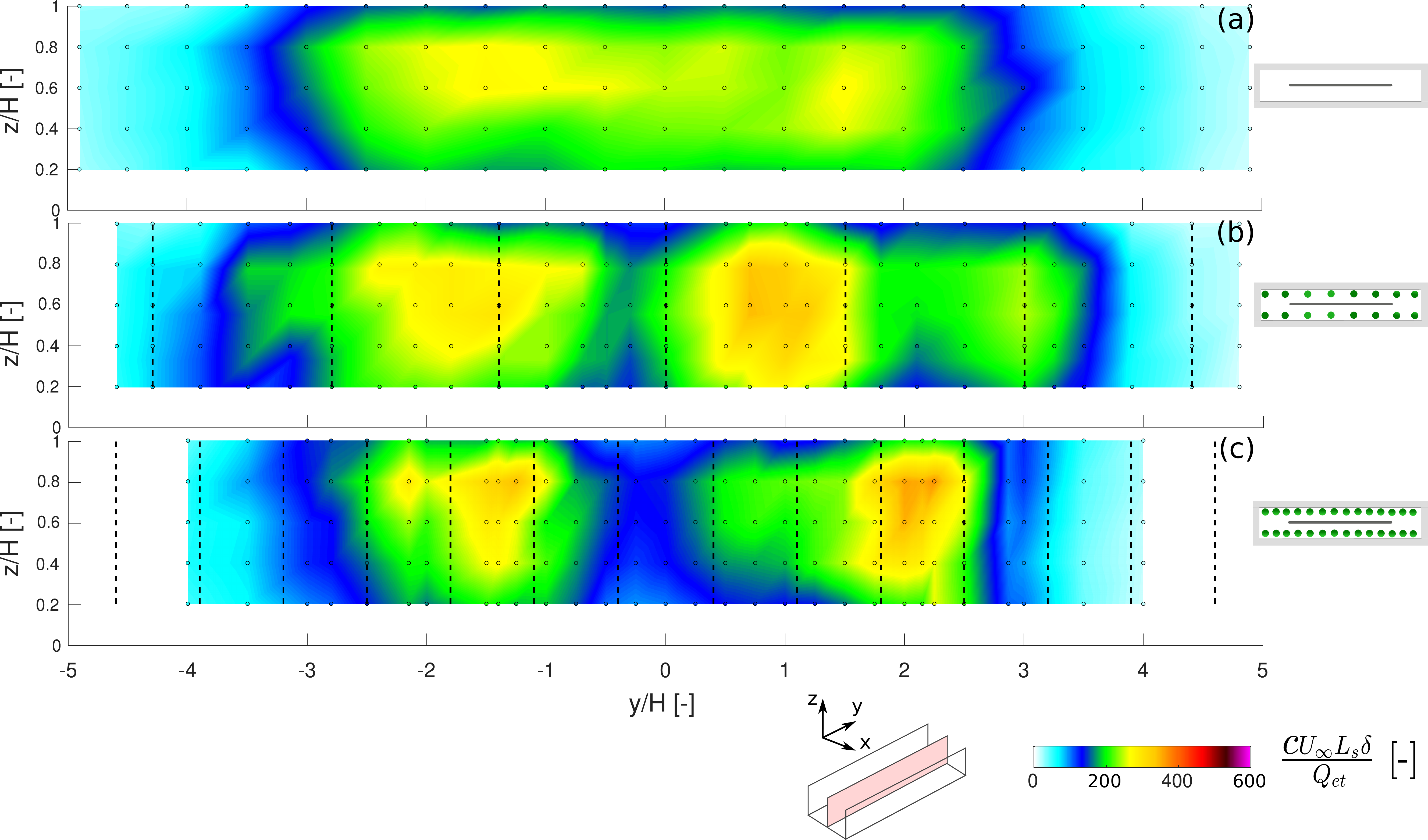}
\caption{Mean concentration of the passive scalar on a vertical section at $x/H$=1. \textit{Zero} (a and d), \textit{Half} (b and e) and \textit{Full} (c and f) configurations are shown. The position of trees is represented by dashed lines.  Measurement points are reported as circles colored according to the measured value.} \label{fig_conc_yz}
\end{figure}

\begin{figure}
\centering
\includegraphics[width=1\columnwidth]{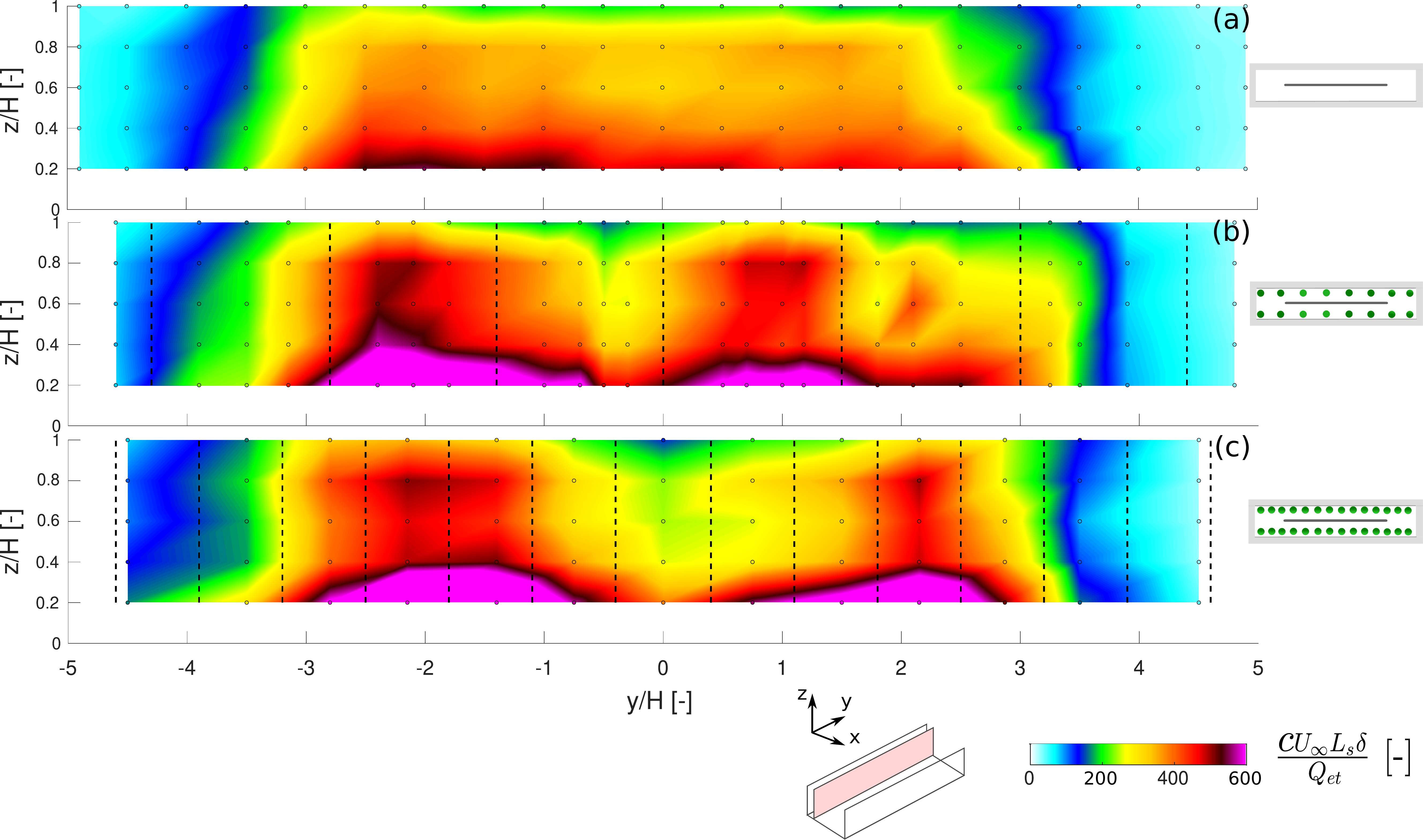}
\caption{Mean concentration of the passive scalar on a vertical section at $x/H$=0.15. \textit{Zero} (a and d), \textit{Half} (b and e) and \textit{Full} (c and f) configurations  are shown. The position of trees is represented by dashed lines.  Measurement points are reported as circles colored according to the measured value.} \label{fig_conc_yz_30}
\end{figure}

To better visualize the spatial distribution of the concentration field, we show in Fig. \ref{fig_conc_yx} an horizontal plane along the canyon axis, at $z/H$=0.6. 
The concentration gradient along the $x$-axis, from the downwind to the upwind wall, is clearly visible in all three configurations. 
As highlighted above, this gradient is enhanced in the vegetated canyons. 
Analysing the concentration near the walls (see also the entire cross sections at $x/H=0.1$ and $x/H=1.9$ in the Supplementary Material), we find that this is due to an average increase in the concentration at the upwind wall in the vegetated canyons, rather than to a decrease of the concentration at the downwind wall, which remains almost constant in the different configurations. 
This result is in line with the studies of Gromke and Ruck \cite{gromke2007influence} and Buccolieri et al. \cite{buccolieri2009aerodynamic} who found that the presence of trees lead to a significant increase in pollutant concentration at the upwind
wall and slight to moderate decrease at the downwind wall.
Figure \ref{fig_conc_yx} also shows that, along the longitudinal axis ($y$-axis), the concentration is almost homogeneous in the \textit{Zero} configuration (panel a), except for the low values at the edges of the domain due to the limited length of the linear source. 
On the other hand, the homogeneity along the $y$-axis is lost when trees are added. 
In the \textit{Full} configuration (panel c), it is possible to identify a region with lower concentration in the middle of the canyon and two nearly symmetric accumulation regions at its sides. 
In the \textit{Half} configuration (panel b), the concentration field is even more heterogeneous and three accumulation regions can be identified.
The same spatial distribution can be inferred from Fig. \ref{fig_conc_yz}, where a vertical section in the middle of the canyon ($x/H$=1) is represented. Again, we observe a homogeneous concentration field in the empty canyon, while pollution peaks are evident in the \textit{Half} and \textit{Full} configurations.
Along the vertical axis, the concentration remains fairly constant. This is typical in the centre of the canyon (i.e. $x/H$=1) and was already visible when focusing on vertical profiles at $x/H$=1 in Fig. \ref{fig_conc_xz}. 
Moving towards the upwind wall (Fig. \ref{fig_conc_yz_30} shows the vertical section at $x/H$=0.15), however, the concentration is greater at street level and a gradient along the vertical axis emerges. 

\begin{figure}
\centering
\includegraphics[width=0.7\columnwidth]{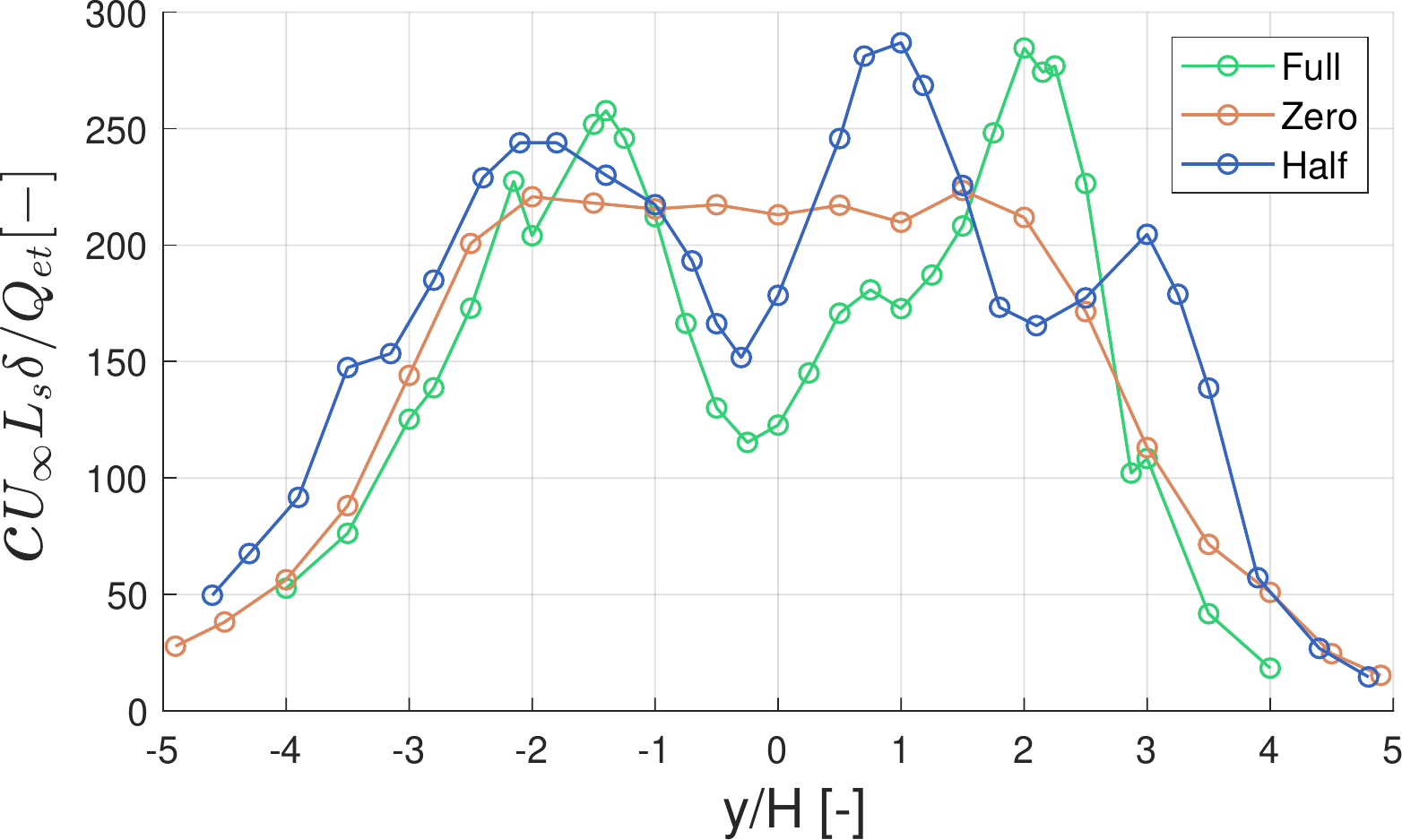}
\caption{Concentration profile along the canyon axis ($x/H=1$) for the three configurations of tree density. Each point is obtained as the average concentration along the vertical axis (z).} \label{fig_concentration_profile}
\end{figure}

By averaging the concentration over the vertical direction ($z$) in the centre of the canyon ($x/H$=1), we obtain in Fig. \ref{fig_concentration_profile} the concentration trend along the longitudinal axis ($y$). 
The curves highlight an homogeneous concentration in the empty canyon and a spatial pattern with pronounced peaks in the vegetated canyons. 
In the \textit{Full} configuration, the peak concentration is approximately twice the minimum concentration measured in the centre of the canyon. 
In Fig. \ref{fig_concentration_profile}, we also observe that the symmetry along the $y$-axis observed for the \textit{Zero} case (orange line) is disturbed in the vegetated configurations (blue and green lines). The reason for this can be found in the irregular and non-identical shape of the trees, and in their random orientation with respect to their axis. 

Finally, from the characterization of the concentration field (Figs. \ref{fig_conc_yx}-\ref{fig_conc_yz_30}), some observations about the local effect of trees on pollution can be deduced. 
For both configurations with trees, the inhomogeneity along the longitudinal axis ($y$) is maintained along the $x$-axis (Fig. \ref{fig_conc_yx}), both near the two rows of trees ($x/H\rightarrow 0$ and $x/H\rightarrow 2$) and in the centre ($x/H=$1). 
Furthermore, the number of concentration peaks (2 and 3 in the \textit{Full} and \textit{Half} configurations, respectively) and their spacing do not correspond to the number and spacing of the trees, represented by dashed lines in Figs. \ref{fig_conc_yx} to \ref{fig_conc_yz_30}.
These two aspects, namely the concentration inhomogeneity along the $y$-axis in the centre of the street and the independence of the concentration peaks on tree pattern, suggest that the variation of the concentration field along the canyon is not due to local effects of trees acting as obstacles. 
Rather, the presence of trees seems to modify the dynamics of flow and dispersion within the whole canyon leading to a different spatial organization of the concentration field at the canyon-scale.

\subsection{Vertical exchange velocity}\label{sec-ud}

While the previous section highlighted the effect of trees on the spatial pattern of the concentration field, in this section we investigate the effect of vegetation on the ventilation efficiency of the canyon.
To this aim, we adopt a box model with one degree of freedom to evaluate the wash-out velocity of the canyon for the different configurations of tree density.

A street canyon can be in general described as a unique box with volume-averaged concentration, $C_{vol}$, and scalar flux through the boundary surfaces at the roof height (of area $L\cdot W$) and at the lateral edges (two rectangles of area $H\cdot W$).
The mass balance for the canyon reads:
\begin{equation}\label{balance_volume}
V \dfrac{\partial C_{vol}}{\partial t}+ \int_{-L/2}^{L/2} \int_0^W  \bigl[\overline{w c} \bigr]_{z = H}  dx dy + \int_0^H \int_0^W  \bigl[\overline{v c} \bigr]_{y =0}^{y=L}  dx dz= Q_{et}, 
\end{equation}
where $V$ is the volume of the canyon, $Q_{et}$ is the mass flow rate of ethane at the source, $\overline{w c}$ and $\overline{v c}$ are the mass fluxes of passive scalar in the vertical and longitudinal directions, respectively. The latter are provided by the time-averaged product of the instantaneous vertical ($w$) or longitudinal velocity ($v$) and concentration $c$. 
In steady state conditions (i.e. $\partial C_{vol}/\partial t=0$) and considering that the lateral boundaries of the volume are solid walls (i.e. $\overline{v c}=0$) in the configuration under study,  the balance for the canyon (Eq. \ref{balance_volume}) becomes:
\begin{equation}\label{balance_volume_2}
\int_{-L/2}^{L/2} \int_0^W  \bigl[\overline{w c} \bigr]_{z = H}  dx dy = Q_{et}.
\end{equation}
Following the study of Soulhac et al. \cite{soulhac2013parametric}, the vertical flux at roof level can be parametrized as the product of a bulk exchange velocity $u_d$ and the difference between the concentration within the canyon ($C_{vol}$) and the concentration in the external flow ($C_{ext}$): 
\begin{equation}\label{ud_def}
u_d  (C_{vol}-C_{ext}) WL= \int_{-L/2}^{L/2} \int_0^W  \bigl[\overline{w c} \bigr]_{z = H}  dy dx.
\end{equation}
Eqs. \ref{balance_volume_2} and \ref{ud_def} can be combined, and the exchange velocity expressed as \cite{salizzoni2009street}:
\begin{equation}\label{ud_def_2}
u_d=\frac{Q_{et}}{C_{vol} WL}.
\end{equation}
or in non-dimensional form as: 
\begin{equation}\label{ud_def_3}
\frac{u_d}{U_\infty}=\frac{L_s \delta}{C_{vol}^* WL},
\end{equation}
where $C_{vol}^*$ is the non-dimensional concentration averaged over the volume and $L_s$ is the source length inside the volume. 
Thanks to this formulation, the vertical exchange velocity can be easily estimated from the quantities measured in the experiment: the flow rate at the source $Q_{et}$ is imposed and monitored by a mass-flow rate controller, while the FID measurements inside the street canyon provide the average concentration in the entire volume ($C_{vol}^*$). 

In this regard, we recall that the choice of reproducing a canyon closed at the lateral edges was made to simplify the estimate of the exchange velocity $u_d$. 
Otherwise, in the case of a canyon with lateral street intersections, the mass balance in Eq. \ref{balance_volume} should be used instead of Eq. \ref{balance_volume_2}. 
As a consequence of that, the flux of passive scalar along the $y-$direction would also appear in Eqs. \ref{ud_def_2} and \ref{ud_def_3} and thus the estimate of $u_d$ would require additional coupled measurements of velocity and concentration. 
In the case of an infinitely long canyon (i.e. a canyon long as the wind-tunnel width), the balance would be applied to a reference volume since the high resolution measurement of the concentration over the entire length of the canyon would be experimentally unfeasible. 
Also in this case, the estimate of the mass flux $\overline{v c}$ at the lateral boundaries of the reference volume would be necessary to compute $u_d$. 

The mean concentration within the canyon was estimated with different spatial averaging techniques of the point measurements: (i) the arithmetic mean of the data, (ii) the mean weighted by the volume associated to each measurement point, and (iii) the mean over a regular grid obtained from interpolation on the concentration data. 
Table \ref{tab-ud} reports the values of $C_{vol}^*$ obtained with these different methods. 
The greatest differences in the estimate of $C_{vol}^*$ are found for the \textit{Full} configuration, where the measurement grid is more irregular due to the presence of trees. 
In this case, the arithmetic mean is expected not to be a good indicator as it does not take into account the uneven spacing of the measuring points.  This limitation is overcome by the weighted average but its value can vary considerably depending on where the measurements are taken with respect to the spatial variability of the concentration field. 
Moreover, the choice of the weights could be not unique and involve a certain degree of arbitrariness affecting the final value. 
The interpolated concentration field can describe the spatial variations by reconstructing (linearly) the concentration pattern even where the measurements are not available. 
In this case, it is important that the interpolation grid is sufficiently fine, at least as refined as the measurement grid.
Regardless the method used, the average level of pollution does not show a trend with the number of trees, and the variation in $C_{vol}^*$ between the configurations (up to 22$\%$) is much lower than the spatial variations induced by the presence of vegetation (Figs. \ref{fig_conc_xz}-\ref{fig_conc_yz_30}).
We also observe that the mean concentration in the \textit{Half} configuration is higher than in the \textit{Full} configuration, where the number of trees is double. This is a remarkable and unexpected result, as it is generally believed that trees, acting as aerodynamic obstacles, are responsible for the accumulation of pollutants in the street. 

To better quantify the effect of trees on street canyon ventilation, we use Eq. \ref{ud_def_3} to estimate the exchange velocity $u_d$, starting from $C_{vol}^*$.
The values of $u_d$ are also presented in Table \ref{tab-ud} and show, once again, that a trend of the ventilation efficiency with the vegetation density is absent, being the \textit{Half} configuration the one exhibiting the lowest exchange rate.
We also note that the estimated values of $u_d$ are higher with respect to those presented by Salizzoni et al. \cite{salizzoni2009street}, Soulhac et al. \cite{soulhac2013parametric}, and Fellini et al. \cite{fellini2020street} for a square cavity ($H/W$=1), confirming that the enlargement of the cavity enhances canyon ventilation.

\begin{figure}
\centering
\includegraphics[width=1\columnwidth]{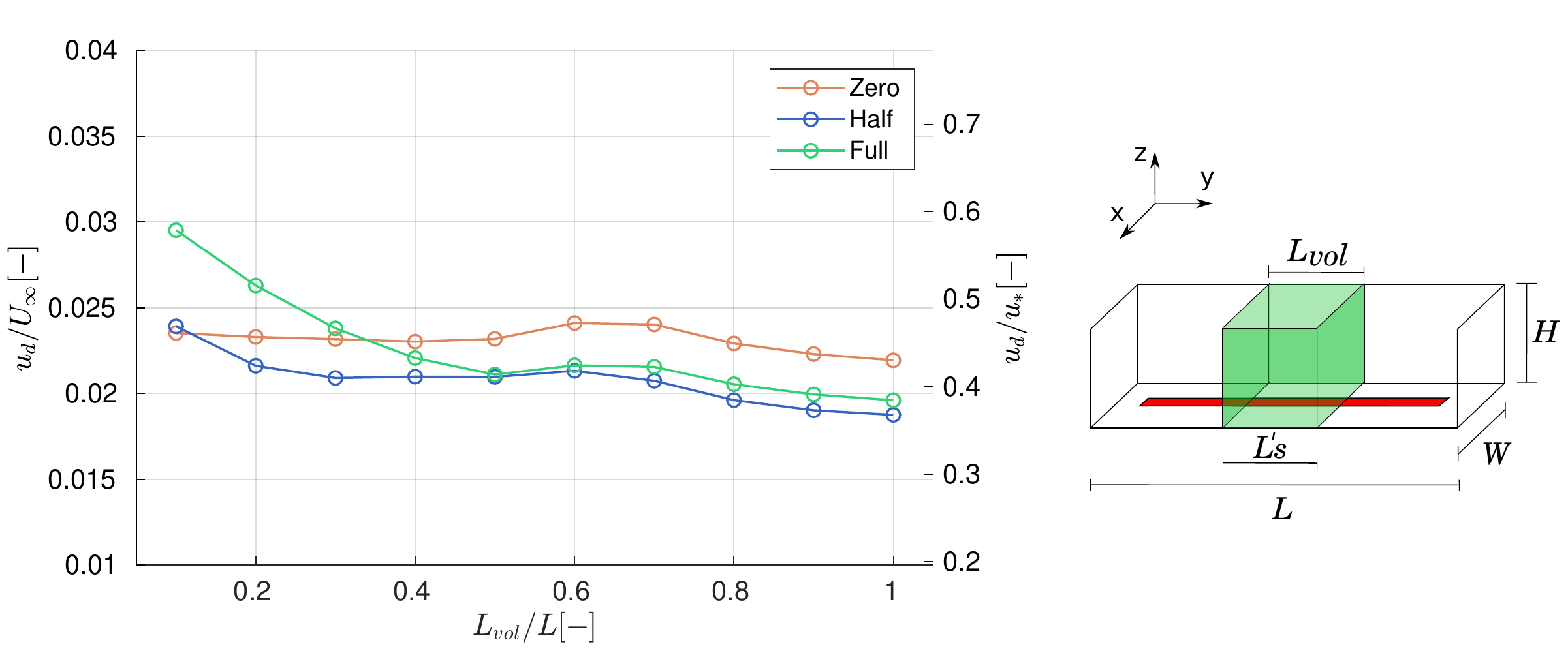}
\caption{Estimate of the exchange velocity $u_d$ as a function of the length ($L_{vol}$) of the reference volume along the longitudinal axis of the canyon.}\label{fig_ud}
\end{figure}

The values reported in Table \ref{tab-ud} have been estimated considering the entire canyon as the reference volume.
In Fig. \ref{fig_ud}, we show how the estimate of $u_d$ varies as a function of the size of the reference volume. 
To this aim, we consider a reference volume centred at $y$=0, extended to the entire width ($W$) and height ($H$) of the canyon, but of variable length ($L_{vol}$) along the longitudinal axis ($y$). 
For the estimate of $u_d$ by means of Eq. \ref{ud_def_3}, the average concentration is computed as the mean of the measurement data interpolated over a regular grid (Table \ref{tab-ud}) inside the reference volume, the source length $L_s$ becomes the effective length included inside the reference volume ($L_s'$ in Fig. \ref{fig_ud}) and the length $L$ is replaced by $L_{vol}$.
For the empty canyon, the exchange rate is almost unchanged as the reference volume increases. This is due to the homogeneity of the concentration field along $y$. In the \textit{Full} configuration, the velocity $u_d$ is greatly overestimated if a volume less than 40\% of the total canyon is considered. 
The reason for this is that, according to the balance in Eq. \ref{ud_def_3}, the lower concentration in the centre of the canyon results in a misleading high ventilation efficiency.
The \textit{Half} configuration shows an intermediate behaviour between the two.
This analysis highlights the importance of a characterization of the concentration field on an extended volume for a correct evaluation of the overall ventilation efficiency. To analyse the effect of the presence of trees in the street, characterizing a single two-dimensional section or a limited volume in the middle of the canyon may lead to misleading conclusions.

\begin{table}[]
\centering
\small
\begin{tabular}{ccccc}
\hline
                                                             &                                & Arithmetic mean  & Weighted mean & Mean of interpolated data \\ \hline
\multicolumn{1}{c|}{\multirow{3}{*}{$C_{vol}^*$ {[}-{]}}}          & \textit{Zero} & 154.84 & 161.32   & 162.96        \\
\multicolumn{1}{c|}{}                                        & \textit{Half} & 197.40 & 185.56   & 190.61        \\
\multicolumn{1}{c|}{}                                        & \textit{Full} & 179.54 & 147.86   & 182.32        \\ \hline
\multicolumn{1}{c|}{\multirow{3}{*}{$u_d/U_\infty$ {[}-{]}}} & \textit{Zero} & 0.023  & 0.022    & 0.022         \\
\multicolumn{1}{c|}{}                                        & \textit{Half} & 0.018  & 0.019    & 0.019         \\
\multicolumn{1}{c|}{}                                        & \textit{Full} & 0.020  & 0.024    & 0.020  \\
\hline      
\end{tabular}
\caption{Estimate of the volume-averaged (normalised) concentration $C_{vol}^*$ for different spatial averaging techniques of the concentration inside the canyon: (i) arithmetic mean, (ii) mean weighted on the reference volume of each measurement point, and (iii) mean of the three-dimensional interpolated concentration field. The corresponding values of the vertical exchange velocity $u_d/U_\infty$, calculated by means of Eq. \ref{ud_def_3}, are also reported. }\label{tab-ud}
\end{table}

\section{Discussion about the effect of lateral boundaries}\label{sec-extrem}

As mentioned above, the canyon geometry with closed lateral borders was adopted to ensure a correct and straightforward estimate of the ventilation efficiency. 
However, this geometry is unusual compared to classic experimental investigations (and realistic urban geometries) and could raise the question if the insertion of side walls radically alters the flow field, and therefore the dispersion process, in the canyon. 
For this reason, in this section we investigate whether the interesting (and partially counter-intuitive) effects of trees on the concentration field, observed in Section \ref{sec-conc}, may have been forced by the presence of the side walls.
To do this, we compare the results presented in the previous sections with the concentration measurements performed in identical experimental conditions, but with different geometries at the lateral borders.
We note that these measurements were carried out on a coarser grid with respect to that adopted for the investigation of the closed canyon. 
While these measurements provide a qualitative description, a deep analysis of the influence of lateral boundaries on the concentration field goes beyond the scope of this paper.

\begin{figure}
\centering
\includegraphics[width=1\columnwidth]{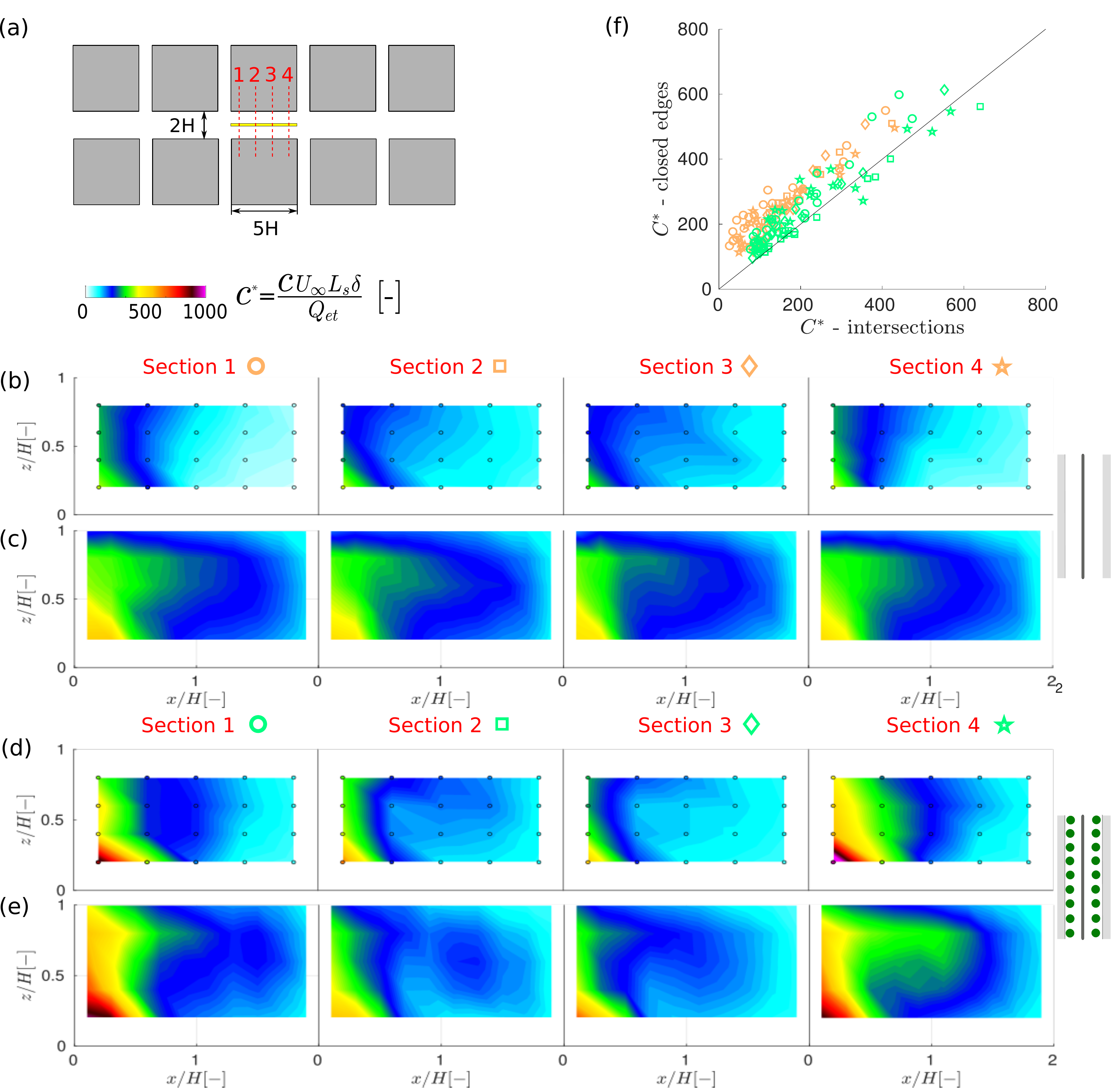}
\caption{(a) Canyon with street intersections at the lateral edges. Concentration in 4 cross-sections for the configuration without trees (b) and with dense trees (d). Comparison with the concentration in the closed cavity, in the \textit{Zero} (c) and \textit{Full} (e) configurations. (f) Comparison of point concentration in the two geometries for the different sections (different markers) and for the different tree densities (different colours).} \label{fig_ale}
\end{figure}

In the first laterally-open geometry \cite{fellini2021modelling}, the reference canyon is part of the network of streets that reproduces the idealized urban district. 
The width ($W=0.2$ m) and height ($H=0.1$ m) of the canyon are the same as in the laterally closed geometry, while its length is 0.5 m as the canyon is limited laterally by two street intersections (Fig. \ref{fig_ale}).
As a consequence, the length-to-width and length-to-height ratios are half compared to the laterally closed geometry and equal to $L/W=2.5$ and $L/H=5$, respectively.
Although the length-to-width ratio is lower and a regular array of building is here present, the street geometry is similar to that adopted by Gromke and Ruck \cite{gromke2007influence,gromke2009impact}, where a laterally open canyon is reproduced.
Concentration measurements were performed over 4 cross-sections (sections 1 to 4) placed at $y/H=\pm1.88$ and $y/H=\pm 0.63$. Results for an empty canyon and a canyon with a full density of trees are presented in panels b and d of Fig. \ref{fig_ale}, respectively. 
The density of trees corresponds to the \textit{Full} configuration presented in Section \ref{sec-wtsetup}. 
Panels c and e refer to the canyon with closed lateral edges (object of study in the previous sections) and report the concentration in the $y/H$ positions corresponding to sections 1 to 4.
As the measurements with lateral walls were not carried out in the exact same positions, we adopted the concentration values obtained from the spatial (linear) interpolation of the point measurements.
Without trees, pollution levels are significantly lower in the open canyon (panel b) with respect to the laterally closed one (panel c).  
This is an expected result as the formation of corner eddies near the intersections provide additional turbulent and advective exchange that favours the canyon ventilation. 
Despite this variation in the pollution levels, we observe that the concentration in panel b remains almost unchanged along the longitudinal axis. This suggests that the concentration field (up to $y/H=\pm1.88$, i.e. sections 1 and 4) in the empty canyon with street intersections is homogeneous along the $y-axis$, as found for the empty canyon with laterally closed edges (panel c).  
Also in the vegetated canyon, the concentration tends to be lower in the canyon with lateral intersections (panel d) with respect to the laterally closed canyon (panel e). We also observe that the concentration trend along $y$ is similar for panels d and e, with sections 1 and 4 exhibiting an increase in the concentration with respect to section 2 and 3. 
This trend is in line with the results found in Section \ref{sec-conc} and suggests that trees induce a three-dimensional concentration field.
The scatterplot in Fig. \ref{fig_ale}.f compares the concentration measurements in the two different geometries (closed edges and lateral street intersections) for the different sections (different markers) and for the different tree densities (different colours). 
This figure points out that the concentration is higher when the lateral ends are closed. This is especially true for the non-vegetated case, where point concentrations increase by over 100\%.
Moreover, the scatterplot highlights a good correlation between the concentration values, with the points that, despite the expected translation along the $y$-axis, follow a linear trend. 
This unexpected good correlation of point concentration in the most part of the domain of two significantly different geometries may be due to the shape of the street intersections and the pattern of the urban canopy: the decelerated longitudinal wind along the narrow ($H/W$ = 1) street canyons generates weak corner eddies which therefore have limited effect on the flow and concentration fields inside the canyon.

\begin{figure}
\centering
\includegraphics[width=1\columnwidth]{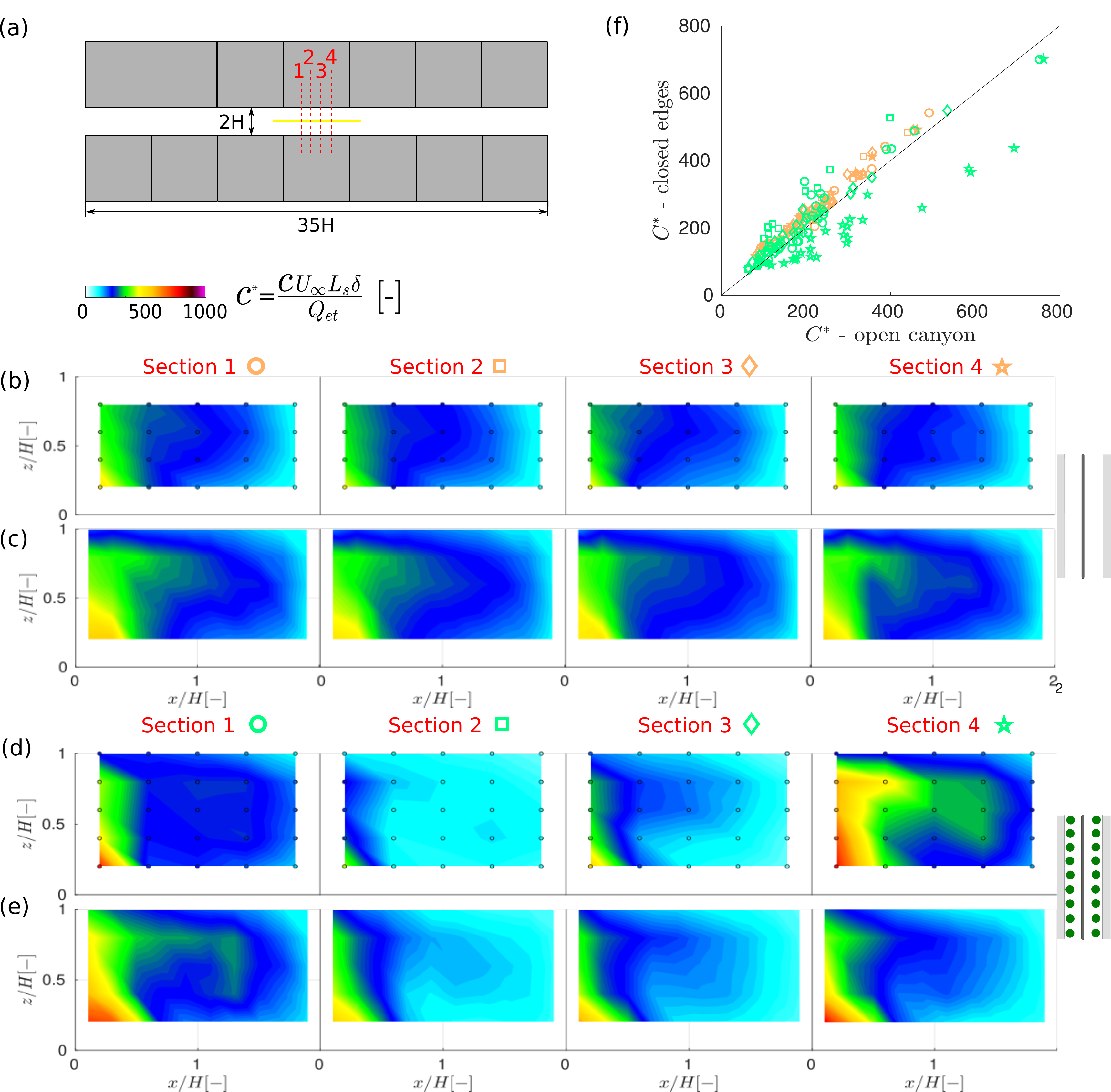}
\caption{(a) Canyon extended to the width of the wind tunnel. Concentration in 4 cross-sections for the configuration without trees (b) and with dense trees (d). Comparison with the concentration in the closed cavity, in the \textit{Zero} (c) and \textit{Full} (e) configurations. (f) Comparison of point concentration in the two geometries for the different sections (different markers) and for the different tree densities (different colours).} \label{fig_giac}
\end{figure}

A similar comparison is performed for the geometry presented in Fig. \ref{fig_giac}.a. 
In this case, the canyon is extended to the entire width of the wind tunnel ($L/H = 35$ and $L/W=17.5$). 
The other geometrical properties of the canyon, as well as the experimental conditions, are the same as presented in Section \ref{sec-wtsetup}.
Concentration measurements are available in four cross-sections placed at $y/H=\pm1.07$ and $y/H=\pm 0.36$. As for Fig. \ref{fig_ale}, we compare the concentration measurements in the four sections (Fig. \ref{fig_giac}.b and d) with the interpolated concentration in the same positions for the geometry with closed lateral edges (Fig. \ref{fig_giac}.c and e). 
Also for the laterally open canyon (panel a), the concentration field shows homogeneity along the $y-axis$ in the absence of trees (panel b), while in the \textit{Full} configuration (panel d) the concentration tends to increase in sections 1 and 4 which are located near the two peaks evident in Fig. \ref{fig_concentration_profile}. 
The comparison between panels b-c together with the scatter plot in Fig. \ref{fig_giac}.f highlight a good agreement between point concentration in the two different geometries (laterally closed and open canyons) for the case without vegetation. 
This suggests that the lateral walls have a negligible effect on the flow field and dispersion in the canyon, at least up to $y/H=\pm1.07$.
Lower agreement is found in the vegetated canyon and especially in section 4. 
As already highlighted in the previous section, the concentration field is strongly affected by the trees that are heterogeneous in shape and placed with random orientation in the canyon. 
These features can therefore lead to slight asymmetries in the concentration field.
The observed differences are more pronounced when comparing point-to-point values in areas with a high concentration gradient, such as in section 4 (located near a peak).

The comparisons discussed in this section suggest that, for the analysed different geometries of lateral ends, the presence of trees in the street induces the inhomogeneity of the concentration field along the $y-$axis. 
The closure of the lateral ends with respect to the laterally open canyon (Fig. \ref{fig_giac}) has negligible effect on the concentration pattern in the case of the empty canyon, while moderate to significant differences in the vegetated canyon may be induced by the shape and orientation of the trees.  
The geometry with lateral intersections, on the other hand, presents more marked differences but the presence of trees seems, even in this case, to trigger the transition from a two-dimensional to a three-dimensional concentration field.

\section{Conclusions}\label{sec-concl}

By means of a high-resolution experimental investigation, we have analysed the effect of trees on ventilation efficiency and pollutant dispersion in an urban street canyon. 

The results show that the concentration field is homogeneous along the longitudinal
axis of the canyon in the \textit{Zero} configuration, and can be considered nearly two-dimensional.
On the other hand, the vegetated streets present a remarkable three-dimensional spatial distribution of the concentration field. 
In particular, concentration peaks alternating with low polluted regions are observed along the canyon both in the \textit{Half} and \textit{Full} configurations and especially in the lower part of the canyon, suggesting high spatial variation in pedestrian exposure.  
Moreover, the presence of trees lead to a significant increase of pollutant concentration at the upwind wall of the street, while the average concentration at the downwind wall is almost constant in the different configurations. 
Despite the great influence of trees on the spatial pattern of pollution, the average concentration in the entire volume of the canyon does not show similar significant variations. Additionally, the \textit{Half} configuration has the highest volume averaged concentration, suggesting that a trend with the tree density is absent.  

The ventilation efficiency of the canyon was quantified by estimating the bulk transfer velocity between the canyon and the boundary layer aloft. 
To this aim, a mass balance within the laterally closed canyon was performed. 
We demonstrated that the characterization of the concentration field over an extended reference volume is necessary to accurately estimate the transfer velocity when trees are present.

Finally, we investigated whether the effect of vegetation on the concentration field is similar in street canyons with different lateral boundaries. 
As expected, the presence of lateral street intersections increases the ventilation of the canyon and therefore decreases the pollution levels. 
However, also in this geometry, the presence of trees in the canyon seems to induce inhomogeneities of the concentration field along the canyon axis. 
In the case of a canyon that extends to the entire width of the wind tunnel, the concentration pattern is quite similar to that observed in the configuration with closed ends. 
This comparison suggests that the results of this study can be generalized to the case of an infinitely long street canyon, which is a classical geometry adopted in both experimental and numerical studies.

To conclude, despite a detailed description of the concentration field at the pedestrian level was not carried out and trees hinder the measures for a complete street canyon characterization, we showed that trees have a non-trivial effect on the spatial distribution of the pollutant concentration, leading to a highly inhomogeneous scalar field and strong pollution gradients in the bottom part of the canyon. 
This suggests that the presence of trees affects the turbulent flow field within the canyon. 
A new measurement campaign is currently underway to investigate the reasons for the formation of pollution peaks in the presence of trees. 

Moreover, we found that the presence of two rows of trees has moderate (up to $20\%$) effect on the vertical exchange velocity. 
On the one hand, this suggests that the estimated value of the vertical exchange velocity could be included, with good approximation, as a constant in parametric models that simulate pollutant dispersion in cities \cite{mchugh1997adms,soulhac2011model,fellini2019propagation}. 
On the other hand, the strong spatial gradients of pollution level due to the presence of trees put into question the reliability of the description of the street canyon as a single box with homogeneous concentration, especially when the focus is on pedestrian exposure. 

Finally, the huge dataset provided by this experimental study can be of great use for the validation of numerical simulations. To this end, the characterization of the boundary layer and of the aerodynamic properties of trees (described in the Section \ref{sec-sim}) are fundamental information. 

\section*{Data availability}

The experimental dataset is available on the website: \url{https://github.com/sfellini/Tree_alpha90_HW05.git}. 
We provide the concentration data inside the canyon and the characterization of the flow field above the buildings.

\section*{Aknowledgements}
We would like to express our gratitude to Horacio Correia for the technical support
in performing the wind tunnel experiments, and to Alessandro De Giovanni, Giacomo Balestrieri, and Lucas Mechinaud for taking part in the experimental campaign. 



\begin{thebibliography}{10}
\expandafter\ifx\csname url\endcsname\relax
  \def\url#1{\texttt{#1}}\fi
\expandafter\ifx\csname urlprefix\endcsname\relax\def\urlprefix{URL }\fi
\expandafter\ifx\csname href\endcsname\relax
  \def\href#1#2{#2} \def\path#1{#1}\fi

\bibitem{bozovic2017blue}
R.~Bozovic, C.~Maksimovic, A.~Mijic, K.~Smith, I.~Suter, M.~Van~Reeuwijk, Blue
  Green Solutions. A Systems Approach to Sustainable and Cost-Effective Urban
  Development, 2017.

\bibitem{oliveira2011cooling}
S.~Oliveira, H.~Andrade, T.~Vaz, The cooling effect of green spaces as a
  contribution to the mitigation of urban heat: A case study in lisbon,
  Building and environment 46~(11) (2011) 2186--2194.

\bibitem{georgakis2017determination}
C.~Georgakis, M.~Santamouris, Determination of the surface and canopy urban
  heat island in athens central zone using advanced monitoring, Climate 5~(4)
  (2017) 97.

\bibitem{litschke2008reduction}
T.~Litschke, W.~Kuttler, On the reduction of urban particle concentration by
  vegetation--a review, Meteorologische Zeitschrift 17~(3) (2008) 229--240.

\bibitem{livesley2016urban}
S.~Livesley, E.~G. McPherson, C.~Calfapietra, The urban forest and ecosystem
  services: impacts on urban water, heat, and pollution cycles at the tree,
  street, and city scale, Journal of environmental quality 45~(1) (2016)
  119--124.

\bibitem{revelli2022green}
F.~Busca, R.~Revelli, Green areas and climate change adaptation in a urban
  environment: The case study of “le vallere” park (turin, italy),
  Sustainability 14~(13) (2022) 8091.

\bibitem{janhall2015review}
S.~Janh{\"a}ll, Review on urban vegetation and particle air
  pollution--deposition and dispersion, Atmospheric environment 105 (2015)
  130--137.

\bibitem{abhijith2017air}
K.~Abhijith, P.~Kumar, J.~Gallagher, A.~McNabola, R.~Baldauf, F.~Pilla,
  B.~Broderick, S.~Di~Sabatino, B.~Pulvirenti, Air pollution abatement
  performances of green infrastructure in open road and built-up street canyon
  environments--a review, Atmospheric Environment 162 (2017) 71--86.

\bibitem{soulhac2008flow}
L.~Soulhac, R.~J. Perkins, P.~Salizzoni, Flow in a street canyon for any
  external wind direction, Boundary-Layer Meteorology 126~(3) (2008) 365--388.

\bibitem{buccolieri2022obstacles}
R.~Buccolieri, O.~S. Carlo, E.~Rivas, J.~L. Santiago, P.~Salizzoni, M.~S.
  Siddiqui, Obstacles influence on existing urban canyon ventilation and air
  pollutant concentration: A review of potential measures, Building and
  Environment (2022) 108905.

\bibitem{allegrini2013wind}
J.~Allegrini, V.~Dorer, J.~Carmeliet, Wind tunnel measurements of buoyant flows
  in street canyons, Building and Environment 59 (2013) 315--326.

\bibitem{murena2016effect}
F.~Murena, B.~Mele, Effect of balconies on air quality in deep street canyons,
  Atmospheric Pollution Research 7~(6) (2016) 1004--1012.

\bibitem{marucci2019effect}
D.~Marucci, M.~Carpentieri, Effect of local and upwind stratification on flow
  and dispersion inside and above a bi-dimensional street canyon, Building and
  Environment 156 (2019) 74--88.

\bibitem{fellini2020street}
S.~Fellini, L.~Ridolfi, P.~Salizzoni, Street canyon ventilation: Combined
  effect of cross-section geometry and wall heating, Quarterly Journal of the
  Royal Meteorological Society (2020).

\bibitem{li2022impact}
H.~Li, Y.~Zhao, B.~S{\"u}tzl, A.~Kubilay, J.~Carmeliet, Impact of green walls
  on ventilation and heat removal from street canyons: Coupling of thermal and
  aerodynamic resistance, Building and Environment 214 (2022) 108945.

\bibitem{gromke2016influence}
C.~Gromke, N.~Jamarkattel, B.~Ruck, Influence of roadside hedgerows on air
  quality in urban street canyons, Atmospheric environment 139 (2016) 75--86.

\bibitem{vos2013improving}
P.~E. Vos, B.~Maiheu, J.~Vankerkom, S.~Janssen, Improving local air quality in
  cities: to tree or not to tree?, Environmental pollution 183 (2013) 113--122.

\bibitem{huang2019impacts}
Y.-d. Huang, M.-z. Li, S.-q. Ren, M.-j. Wang, P.-y. Cui, Impacts of
  tree-planting pattern and trunk height on the airflow and pollutant
  dispersion inside a street canyon, Building and Environment 165 (2019)
  106385.

\bibitem{gromke2007influence}
C.~Gromke, B.~Ruck, Influence of trees on the dispersion of pollutants in an
  urban street canyon - {Experimental} investigation of the flow and
  concentration field, Atmospheric Environment 41~(16) (2007) 3287--3302.

\bibitem{gromke2009impact}
C.~Gromke, B.~Ruck, On the impact of trees on dispersion processes of traffic
  emissions in street canyons, Boundary-Layer Meteorology 131~(1) (2009)
  19--34.

\bibitem{gromke2012pollutant}
C.~Gromke, B.~Ruck, Pollutant concentrations in street canyons of different
  aspect ratio with avenues of trees for various wind directions,
  Boundary-Layer Meteorology 144~(1) (2012) 41--64.

\bibitem{buccolieri2009aerodynamic}
R.~Buccolieri, C.~Gromke, S.~Di~Sabatino, B.~Ruck, Aerodynamic effects of trees
  on pollutant concentration in street canyons, Science of the Total
  Environment 407~(19) (2009) 5247--5256.

\bibitem{buccolieri2011analysis}
R.~Buccolieri, S.~M. Salim, L.~S. Leo, S.~Di~Sabatino, A.~Chan, P.~Ielpo,
  G.~de~Gennaro, C.~Gromke, Analysis of local scale tree--atmosphere
  interaction on pollutant concentration in idealized street canyons and
  application to a real urban junction, Atmospheric Environment 45~(9) (2011)
  1702--1713.

\bibitem{gromke2008dispersion}
C.~Gromke, R.~Buccolieri, S.~Di~Sabatino, B.~Ruck, Dispersion study in a street
  canyon with tree planting by means of wind tunnel and numerical
  investigations--evaluation of cfd data with experimental data, Atmospheric
  Environment 42~(37) (2008) 8640--8650.

\bibitem{gromke2015influence}
C.~Gromke, B.~Blocken, Influence of avenue-trees on air quality at the urban
  neighborhood scale. part i: Quality assurance studies and turbulent schmidt
  number analysis for rans cfd simulations, Environmental Pollution 196 (2015)
  214--223.

\bibitem{vranckx2015impact}
S.~Vranckx, P.~Vos, B.~Maiheu, S.~Janssen, Impact of trees on pollutant
  dispersion in street canyons: A numerical study of the annual average effects
  in antwerp, belgium, Science of the Total Environment 532 (2015) 474--483.

\bibitem{salim2011numerical}
S.~M. Salim, S.~C. Cheah, A.~Chan, Numerical simulation of dispersion in urban
  street canyons with avenue-like tree plantings: comparison between rans and
  les, Building and Environment 46~(9) (2011) 1735--1746.

\bibitem{moonen2013performance}
P.~Moonen, C.~Gromke, V.~Dorer, Performance assessment of large eddy simulation
  (les) for modeling dispersion in an urban street canyon with tree planting,
  Atmospheric environment 75 (2013) 66--76.

\bibitem{merlier2018lattice}
L.~Merlier, J.~Jacob, P.~Sagaut, Lattice-boltzmann large-eddy simulation of
  pollutant dispersion in street canyons including tree planting effects,
  Atmospheric Environment 195 (2018) 89--103.

\bibitem{Irwin1981}
H.~Irwin, The design of spires for wind simulation, Journal of Wind Engineering
  and Industrial Aerodynamics 7~(3) (1981) 361 -- 366.
\newblock \href {https://doi.org/https://doi.org/10.1016/0167-6105(81)90058-1}
  {\path{doi:https://doi.org/10.1016/0167-6105(81)90058-1}}.

\bibitem{soulhac2010dispersion}
L.~Soulhac, P.~Salizzoni, Dispersion in a street canyon for a wind direction
  parallel to the street axis, Journal of Wind Engineering and Industrial
  Aerodynamics 98~(12) (2010) 903--910.

\bibitem{moonen2011evaluation}
P.~Moonen, V.~Dorer, J.~Carmeliet, Evaluation of the ventilation potential of
  courtyards and urban street canyons using rans and les, Journal of Wind
  Engineering and Industrial Aerodynamics 99~(4) (2011) 414--423.

\bibitem{marro2020high}
M.~Marro, H.~Gamel, P.~M{\'e}jean, H.~Correia, L.~Soulhac, P.~Salizzoni,
  High-frequency simultaneous measurements of velocity and concentration within
  turbulent flows in wind-tunnel experiments, Experiments in Fluids 61~(12)
  (2020) 1--13.

\bibitem{meroney1996study}
R.~N. Meroney, M.~Pavageau, S.~Rafailidis, M.~Schatzmann, Study of line source
  characteristics for 2-d physical modelling of pollutant dispersion in street
  canyons, Journal of wind Engineering and industrial Aerodynamics 62~(1)
  (1996) 37--56.

\bibitem{fackrell1980flame}
J.~Fackrell, A flame ionisation detector for measuring fluctuating
  concentration, Journal of Physics E: Scientific Instruments 13~(8) (1980)
  888.

\bibitem{pavageau1999wind}
M.~Pavageau, M.~Schatzmann, Wind tunnel measurements of concentration
  fluctuations in an urban street canyon, Atmospheric Environment 33~(24-25)
  (1999) 3961--3971.

\bibitem{carpentieri2012wind}
M.~Carpentieri, P.~Hayden, A.~G. Robins, Wind tunnel measurements of pollutant
  turbulent fluxes in urban intersections, Atmospheric Environment 46 (2012)
  669--674.

\bibitem{nironi2015dispersion}
C.~Nironi, P.~Salizzoni, M.~Marro, P.~Mejean, N.~Grosjean, L.~Soulhac,
  Dispersion of a passive scalar fluctuating plume in a turbulent boundary
  layer. part i: Velocity and concentration measurements, Boundary-layer
  meteorology 156~(3) (2015) 415--446.

\bibitem{vidali2022wind}
C.~Vidali, M.~Marro, H.~Correia, L.~Gostiaux, S.~Jallais, D.~Houssin,
  E.~Vyazmina, P.~Salizzoni, Wind-tunnel experiments on atmospheric heavy gas
  dispersion: Metrological aspects, Experimental Thermal and Fluid Science 130
  (2022) 110495.

\bibitem{comte1976hot}
G.~Comte-Bellot, Hot-wire anemometry, Annual review of fluid mechanics 8~(1)
  (1976) 209--231.

\bibitem{tritton2012physical}
D.~J. Tritton, Physical fluid dynamics, Springer Science \& Business Media,
  2012.

\bibitem{gromke2008aerodynamic}
C.~Gromke, B.~Ruck, Aerodynamic modelling of trees for small-scale wind tunnel
  studies, Forestry 81~(3) (2008) 243--258.

\bibitem{meroney1968characteristics}
R.~N. Meroney, Characteristics of wind and turbulence in and above model
  forests, Journal of Applied Meteorology 7~(5) (1968) 780--788.

\bibitem{meroney1980wind}
R.~N. Meroney, Wind-tunnel simulation of the flow over hills and complex
  terrain, Journal of Wind Engineering and Industrial Aerodynamics 5~(3-4)
  (1980) 297--321.

\bibitem{chen1995wind}
J.~Chen, T.~Black, M.~Novak, R.~Adams, A wind tunnel study of turbulent airflow
  in forest clearcuts, Wind and trees (1995) 71--87.

\bibitem{stacey1994wind}
G.~Stacey, R.~Belcher, C.~Wood, B.~Gardiner, Wind flows and forces in a model
  spruce forest, Boundary-Layer Meteorology 69~(3) (1994) 311--334.

\bibitem{gromke2011vegetation}
C.~Gromke, A vegetation modeling concept for building and environmental
  aerodynamics wind tunnel tests and its application in pollutant dispersion
  studies, Environmental pollution 159~(8-9) (2011) 2094--2099.

\bibitem{manickathan2018comparative}
L.~Manickathan, T.~Defraeye, J.~Allegrini, D.~Derome, J.~Carmeliet, Comparative
  study of flow field and drag coefficient of model and small natural trees in
  a wind tunnel, Urban Forestry \& Urban Greening 35 (2018) 230--239.

\bibitem{guan2003wind}
D.~Guan, Y.~Zhang, T.~Zhu, A wind-tunnel study of windbreak drag, Agricultural
  and forest meteorology 118~(1-2) (2003) 75--84.

\bibitem{bitog2011wind}
J.~Bitog, I.-B. Lee, H.-S. Hwang, M.-H. Shin, S.-W. Hong, I.-H. Seo,
  E.~Mostafa, Z.~Pang, A wind tunnel study on aerodynamic porosity and
  windbreak drag, Forest Science and technology 7~(1) (2011) 8--16.

\bibitem{lee2014shelter}
J.-P. Lee, E.-J. Lee, S.-J. Lee, Shelter effect of a fir tree with different
  porosities, Journal of Mechanical Science and Technology 28~(2) (2014)
  565--572.

\bibitem{velardeestimation}
J.~G. Velarde, J.~H. Blanco, J.~M. Aliseda, Estimation of optical porosity or
  canopy structure of two species of tree with hemispherical and vertical
  images, WSEAS Transactions on Environment and Development 14~(11) (2018)
  112--124.

\bibitem{perret2019atmospheric}
L.~Perret, J.~Basley, R.~Mathis, T.~Piquet, The atmospheric boundary layer over
  urban-like terrain: influence of the plan density on roughness sublayer
  dynamics, Boundary-Layer Meteorology 170~(2) (2019) 205--234.

\bibitem{castro1977flow}
I.~Castro, A.~Robins, The flow around a surface-mounted cube in uniform and
  turbulent streams, Journal of fluid Mechanics 79~(2) (1977) 307--335.

\bibitem{oke2017urban}
T.~R. Oke, G.~Mills, A.~Christen, J.~A. Voogt, Urban climates, Cambridge
  University Press, 2017.

\bibitem{fisher2006meteorology}
B.~Fisher, J.~Kukkonen, M.~Piringer, M.~Rotach, M.~Schatzmann, Meteorology
  applied to urban air pollution problems: concepts from cost 715, Atmospheric
  Chemistry and Physics 6~(2) (2006) 555--564.

\bibitem{raupach2006momentum}
M.~Raupach, D.~Hughes, H.~Cleugh, Momentum absorption in rough-wall boundary
  layers with sparse roughness elements in random and clustered distributions,
  Boundary-Layer Meteorology 120~(2) (2006) 201--218.

\bibitem{grimmond1999aerodynamic}
C.~Grimmond, T.~R. Oke, Aerodynamic properties of urban areas derived from
  analysis of surface form, Journal of Applied Meteorology and Climatology
  38~(9) (1999) 1262--1292.

\bibitem{garbero2010experimental}
V.~Garbero, P.~Salizzoni, L.~Soulhac, Experimental study of pollutant
  dispersion within a network of streets, Boundary-layer meteorology 136~(3)
  (2010) 457--487.

\bibitem{britter2003flow}
R.~Britter, S.~Hanna, Flow and dispersion in urban areas, Annual review of
  fluid mechanics 35~(1) (2003) 469--496.

\bibitem{hinze1975turbulence}
J.~Hinze, Turbulence, {S}econd edition, McGraw-Hill, New York, 1975.

\bibitem{tennekes1972first}
H.~Tennekes, J.~L. Lumley, J.~L. Lumley, et~al., A first course in turbulence,
  MIT press, 1972.

\bibitem{soulhac2013parametric}
L.~Soulhac, P.~Salizzoni, P.~Mejean, R.~Perkins, Parametric laws to model urban
  pollutant dispersion with a street network approach, Atmospheric Environment
  67 (2013) 229--241.

\bibitem{salizzoni2009street}
P.~Salizzoni, L.~Soulhac, P.~Mejean, Street canyon ventilation and atmospheric
  turbulence, Atmospheric Environment 43~(32) (2009) 5056--5067.

\bibitem{fellini2021modelling}
S.~Fellini, Modelling pollutant dispersion at the city and street scales, Ph.D.
  thesis, Politecnico di Torino \& {\'E}cole Centrale de Lyon (2021).

\bibitem{mchugh1997adms}
C.~McHugh, D.~Carruthers, H.~Edmunds, Adms--urban: an air quality management
  system for traffic, domestic and industrial pollution, International Journal
  of Environment and Pollution 8~(3-6) (1997) 666--674.

\bibitem{soulhac2011model}
L.~Soulhac, P.~Salizzoni, F.-X. Cierco, R.~Perkins, The model {SIRANE} for
  atmospheric urban pollutant dispersion; part {I}, presentation of the model,
  Atmospheric Environment 45~(39) (2011) 7379--7395.

\bibitem{fellini2019propagation}
S.~Fellini, P.~Salizzoni, L.~Soulhac, L.~Ridolfi, Propagation of toxic
  substances in the urban atmosphere: A complex network perspective,
  Atmospheric Environment 198 (2019) 291--301.

\end{thebibliography}

\newpage

\setcounter{figure}{0}
\setcounter{section}{0}

\renewcommand{\thesection}{S\arabic{section}}
\renewcommand{\thefigure}{S\arabic{figure}}

\end{document}